\documentclass[aps,prb,preprint,groupedaddress,showpacs]{revtex4-1}
\usepackage{epsfig}
\usepackage{graphics}
\usepackage{amssymb,amsmath}
\usepackage{color}
\usepackage{siunitx}
\usepackage{float}
\usepackage{multirow}
\usepackage{enumitem}

\renewcommand{\theenumi}{(\Asbuk{enumi})}
\begin{document}

\leftmargin -2cm
\def\choosen{\atopwithdelims..}


 \boldmath

\title{Prediction of new $sp^3$ silicon and germanium allotropes from the
topology-based multiscale method} \unboldmath

\author{\firstname{Vladimir A. }\surname{Saleev}}\email{saleev@samsu.ru}
\author{\firstname{Alexandra V. } \surname{Shipilova}}\email{alexshipilova@samsu.ru}
\affiliation{Department of Physics, Samara National Research
University - 34, Moskovskoe shosse, Samara, 443086, Russian
Federation}
\author{\firstname{Davide M. }\surname{Proserpio}} \email{davide.proserpio@unimi.it}
\affiliation{Universit\`a degli Studi di Milano, Dipartimento di
Chimica - via Golgi, 19, 20133 Milano, Italy} \affiliation{Samara
Center for Theoretical Materials Science (SCTMS), Samara National
Research University - 34, Moskovskoe shosse, Samara, 443086, Russian
Federation}
\author{\firstname{Giuseppe }\surname{Fadda}} \email{gfadda@dmsa.unipd.it}
\affiliation{Research and Education Center for Physics of Open
Nonequilibrium Systems, Samara National Research University - 34,
Moskovskoe shosse, Samara, 443086, Russian Federation}

\begin{abstract}
  This article continues our recent publication \cite{BPSS} in which we have presented a comprehensive computational study of
  $sp^3$  carbon allotropes based on the topologies proposed for zeolites. Here
  we predict six new silicon and six new germanium allotropes
  which have similar group symmetries and topologies as those predicted for
  the early carbon allotropes, and study their structural, elastic, vibrational, electronic, and optical properties.
\end{abstract}

 \maketitle

\section{Introduction}
\label{sec:1} In recent years, carbon allotropes have been the focus
of an intense research activity (see \onlinecite{BPSS} and
references therein); among the methods used to find new potentially
interesting allotropes, many are now well-known to physicists, such
as evolutionary algorithms \cite{JChemPhys.124.244704,Oganov2011}.
The topological method used in Ref.~\onlinecite{BPSS} stems from a
radically different philosophy, namely starting from well-known
structures (instead of looking for new ones), extracting
the connectivity of their fundamental building blocks (see
\onlinecite{BPinOganov2011,CrystGrowthDes.14.3576} and references
therein), and then replacing the latter by any other elementary
structure; for instance in Ref.~\onlinecite{BPSS}, silicates and zeolites
were used as starting structures, their building blocks being
obviously the silicon tetrahedra SiO$_4^{4-}$, the new elementary
structures atoms of carbon (in \onlinecite{BPSS}) or silicon or
germanium (in the present paper). Once done, the newly-found
structures are relaxed; as the number of known zeolites and
silicates is of the order of several hundreds of
thousands~\cite{JPhysChem.113.21353,ZKristallogr.212.768}, some
filtering procedure (described in \onlinecite{BPSS} and
\S~\ref{sec:2}) is necessary.\par Because of the chemical similarity
of Si and Ge with C, as of course all of them belong to Group 14
(IUPAC nomenclature), it seemed a natural idea to extend the
previous study to silicon and germanium. But instead of focusing on
structural and mechanical properties only, we have emphasized here
the generic physical properties of the allotropes (\S~\ref{sec:4}),
including a discussion of the electronic, dielectric, optical, and
vibrational properties.\par
Theoretical prediction of silicon and
germanium allotropes is a subject of great interest because of
their potential use in the semiconductor industry.
Very recent works devoted to the silicon and
germanium allotropes were aimed at finding prospective structures
for photovoltaic applications, as silicon still keeps the
leading position there (see for instance
\onlinecite{RenewSustEnergRev.15.1625,RenewSustEnergRev.15.2165}
and references therein). Several novel metastable phases of $sp^3$-bonded
silicon, obtained through a structural search based on carbon analogues as well
as on random-structure searching approaches, were
predicted~\cite{Zwijnenberg2010,Baburin2012,Haberl2016,Fan2015,He2015,Mujica2015,Botti,NatMater.14.169}.
Their total energy remains within 0.15 eV per atom from diamond silicon, and
the band gaps are located between 1 and 1.5 eV; some of these structures have
shown good absorption properties of solar
light. Here we study 5 previously unknown $sp^3$ allotropes of
silicon and germanium which are within 0.05 eV of the diamond phases, making
them more energetically feasible; the sixth structure investigated here, namely
$\#28$, suggested in our previous work as a carbon allotrope, has already been
studied as a phase of silicon and germanium in Ref.~\onlinecite{Mujica2015}.

Another interesting conclusion of the present work is that
distorted structures induce a severe reduction in $s$-$p$ overlap, which
might even provide for a metallic phase of germanium as discussed recently for
some novel metastable phases of germanium~\cite{Baburin2012}.

\section{Approach for structure searching}
\label{sec:2}

\subsection{Methodology}

\begin{figure}
\includegraphics[width=0.5\textwidth, clip=]{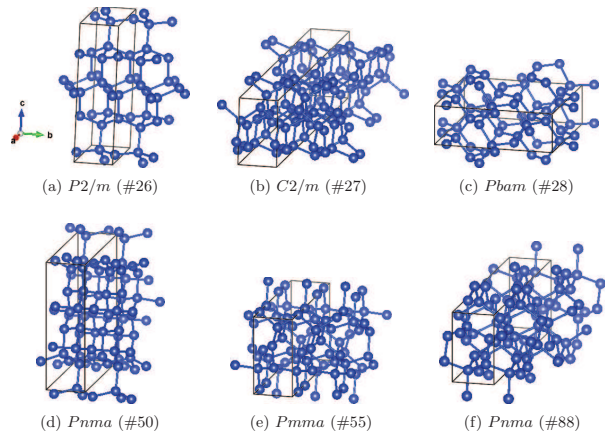}
\caption{Perspective view of the silicon allotropes $\#26$,
$\#27$,$\#28$,$\#50$,$\#55$, and $\#88$.}\label{fig:structure}
\end{figure}

Details are given in \onlinecite{BPSS}; here we
provide the necessary information. The topological analysis of many
of the carbon allotropes found using a variety of methods (see
\onlinecite{BPSS} and references therein) has shown that they are
generally related to silicates or zeolites; the idea was thus to
start from the latter structures (of which hundreds of thousands are
known) and by contracting the oxygen links in the Si--O bonds, to
generate a large number of tetrahedral Si nets. The problem was then
to filter among these structures the ones which could be
interesting: the 3- and 4-ring nets were discarded (in order to
avoid excessive strain in the silicon structures), the rest then submitted
to a geometrical post-screening (mainly based on the nearest and
next-to-nearest distances between Si atoms); progressively more
refined methods (first classical potentials, then tight-binding)
were used to determine the most favorable structures,
and diamond- and lonsdaleite-related nets were rejected on the basis
of their popularity in the literature  \cite{34,35}. The final set
consisted of the six structures with lowest energy $\#26$,
$\#27$, $\#28$, $\#50$, $\#55$, and $\#88$, which are the object of the present
paper. The optimized crystal structures of the predicted silicon allotropes are
shown in Fig.~ \ref{fig:structure}; the germanium allotropes are not shown as
they are very similar.

\subsection{Nomenclature}

The structures studied here, as already discussed above, are the same studied
in Ref.~\onlinecite{BPSS}, and are part of the SACADA (SAmara Carbon Allotrope
DAtabase)~\cite{SACADA1,SACADA2}: see Table~\ref{tab:SACADA} for the
correspondence.
\begin{table}[!htb]
    \centering
    \begin{tabular}{c|c||c|c}
    Present&\multirow{2}{*}{SACADA}&Present&\multirow{2}{*}{SACADA}\\[-3mm]
    work&&work&\\\hline
    \#26&4\^{}8T15&\#27&4\^{}8T16\\
    \#28&4\^{}6T17&\#50&4\^{}7T12\\
    \#55&4\^{}6T16&\#88&4\^{}6T18
    \end{tabular}
    \caption{Correspondence between the labels used in the present paper and
    those used in SACADA for the equivalent carbon allotropes}\label{tab:SACADA}
\end{table}
The labels used in the present paper are the last two digits of the
corresponding hypothetical zeolite structures in the Deem
database~\cite{JPhysChem.113.21353}.

\section{Computational methodology}
\label{sec:3} The main part of the computations, that is all electronic,
dielectric, vibrational, and mechanical properties, plus the Raman shifts and
infrared absorption spectra, was done with the CRYSTAL14 package
\cite{CRYSTAL1,CRYSTAL2}, following the same methodology of
Ref~\onlinecite{BPSS}. The generalized-gradient approximation (GGA) and
the Perdew-Burke-Erzenhof (PBE) \cite{PBE} parameterizations for the
exchange-correlation term were adopted; the basis set used is the
triple-$\zeta$ valence with polarization (TZVP) as developed by Peitinger et
al.  \cite{Basis}. Structural relaxation was done using the conjugate
gradient method; convergence was deemed achieved for forces below 3 meV
${\AA}^{-1}$ and stresses below 0.02 GPa. As to the Brillouin Zone (BZ), the
usual Monkhorst-Pack sampling \cite{MPpoints} was used and tested for
convergence; the Shrinking Factors for the $k$-point generation for the unit
cell and for $2\times 2\times 2$ supercells are $\{8,8,8\}$ and $\{4,4,4\}$
respectively. Raman shift spectra were computed for polycrystalline powder as
implemented in the CRYSTAL14 package.\par
For the computations of the complex refractive indices and
dielectric tensors  in the visible and ultraviolet (UV) regions we used the
VASP package instead \cite{VASP1,VASP2}. A plane-wave basis set with a cutoff
energy of 500~eV and 350~eV for Si and Ge respectively was used throughout the
calculations. For the basic ground-state properties the GGA-PBE was selected
and the integration over the BZ performed using Monkhorst-Pack grids of
$7\times7\times7$ $k$-points. For optical properties, we also used within both
CRYSTAL14 and VASP the screened
Heyd, Scuseria, and Ernzerhof (HSE06)~\cite{HSE06} hybrid functional.\par
In order to ascertain the quality of the present computational approach, we
have performed for the diamond structure a check on all the computed
properties. As can be seen in Table \ref{Table:1}, the adopted method gives a
fair agreement with experiment for the basic properties, including the band gap
and the bulk modulus (two well-known issues of GGA-PBE); as to the computed
optical properties of diamond Si and Ge and the details regarding the adopted
methodology, see \S~\ref{sec:45}.

\section{Results}
\label{sec:4}
\subsection{Energetics of the various allotropes}
\label{sec:41} 
\begin{figure}
\includegraphics[width=1.0\textwidth, clip=]{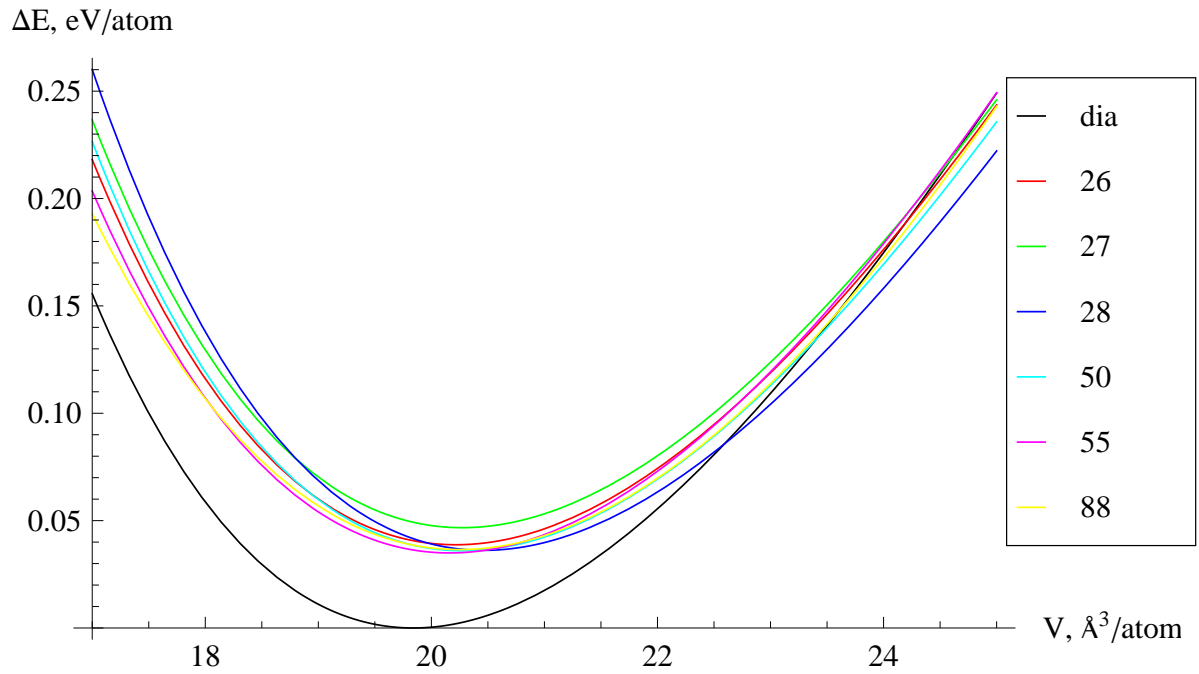}
\caption{Energy per atom as a function of volume for silicon
allotropes.} \label{fig:EVSi}
\end{figure}

\begin{figure}
\includegraphics[width=1.0\textwidth, clip=]{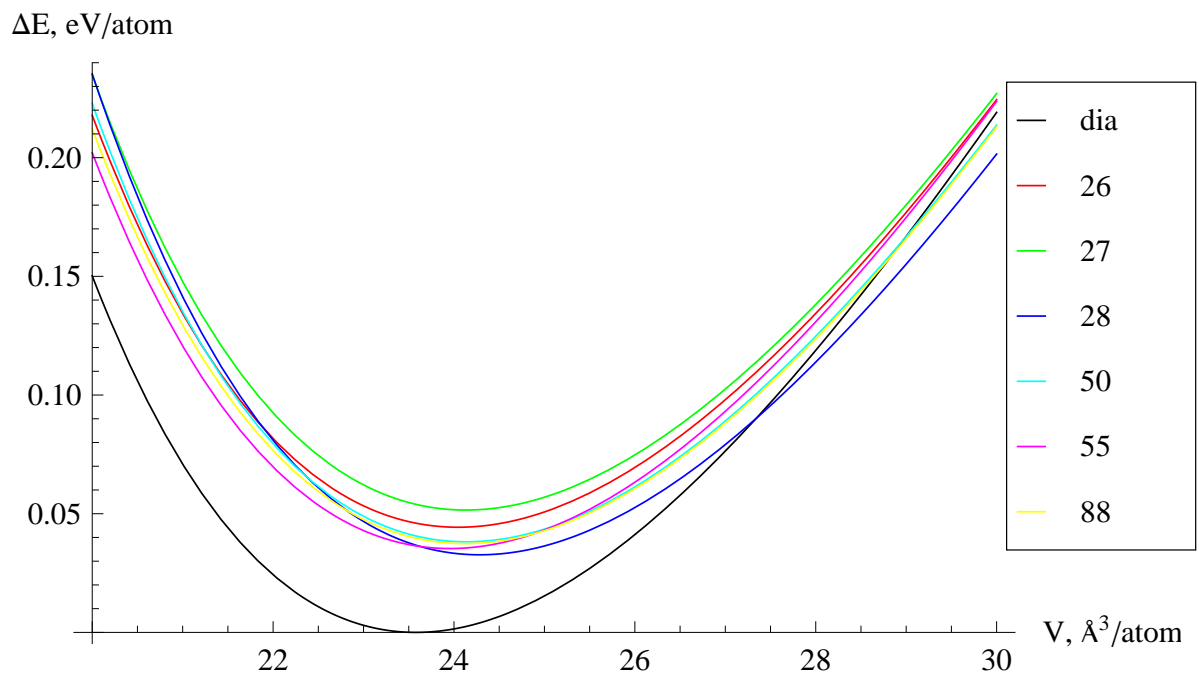}
\caption{Energy per atom as a function of volume for germanium
allotropes.} \label{fig:EVGe}
\end{figure}

\begin{figure}\includegraphics[width=1.0\textwidth, clip=]{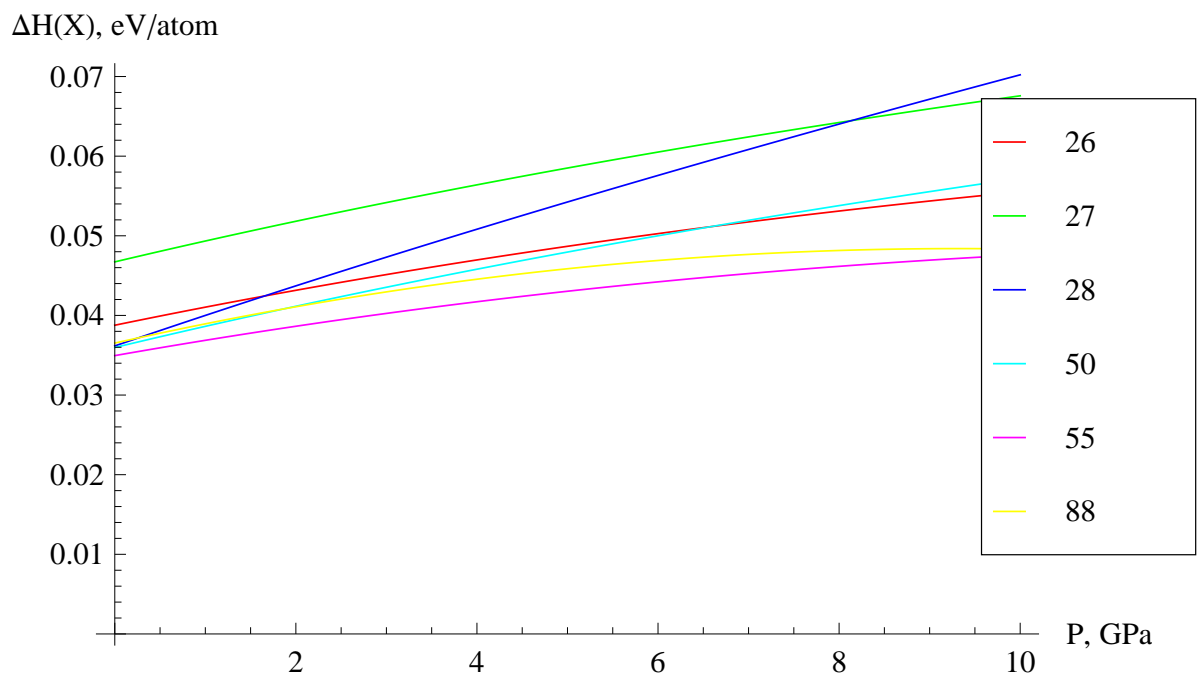}
\caption{Enthalpy difference $\Delta H(P)$ per atom as a function
of pressure for silicon allotropes.}\label{fig:HSi}
\end{figure}

\begin{figure}\includegraphics[width=1.0\textwidth, clip=]{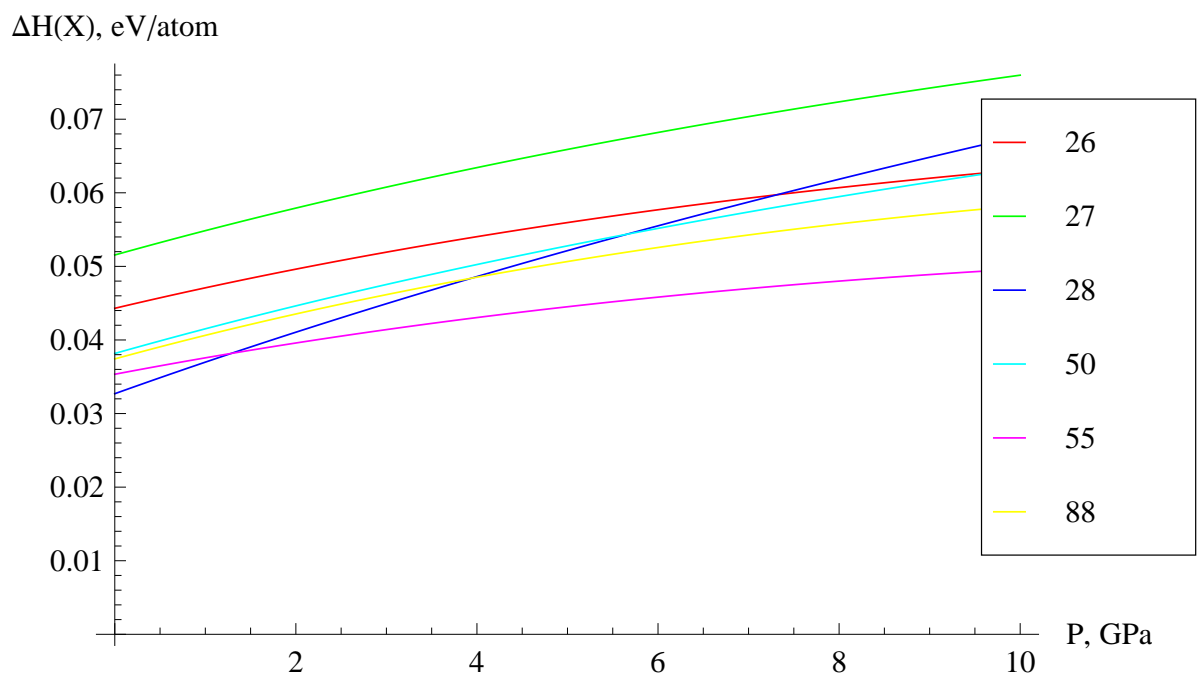}
\caption{Enthalpy difference $\Delta H(P)$ per atom as a function
of pressure for germanium allotropes.}\label{fig:HGe}
\end{figure}

For the basic computations (made at \SI{0}{K}), the
enthalpy is the thermodynamic potential:
\begin{eqnarray}
H(V)&=&E(V)+p(V)\times V.\end{eqnarray} After calculation of the crystal
energy $E$ as a function of the volume $V$ (see Figs. \ref{fig:EVSi} and
\ref{fig:EVGe}), we fit this dependence using the Birch-Murnaghan
equation of state \cite{BM} to find the $p(V)=-\frac{dE}{dV}$ dependence
and the differences of enthalpies  $\Delta H(p)=
H_X(p)-H_{dia}(p)$ between phase~$X$ and diamond as a parametric function of
pressure $p$.
 As seen in Figs. \ref{fig:HSi} and
\ref{fig:HGe}, none of the new allotropes can be considered
high-pressure variants, exactly as for carbon: indeed the difference
with the diamond structure increases with pressure.

At zero pressure and non-zero temperature $T$, the Gibbs
free energy is the thermodynamic potential: \begin{equation}
F(T)=E(0)+E(T)-T\times S(T),
\end{equation}
where $E(0)$ is the crystal energy at $T=0$, $E(T)$ the
thermal energy and $S(T)$ the entropy. $E(T)$ and $S(T)$ can
be calculated in the quasi-harmonic approximation (see
\onlinecite{thermodynamics} for details of implementation in the CRYSTAL14
package). To
calculate the entropy and thermal energy as a function of temperature, the
phonon frequency spectrum of crystal has to be calculated:
\begin{eqnarray}
S(T)&=& k_B T \left(\frac{\partial \log Q}{\partial T}\right)_{V,N}+k_B\log Q,\\
E(T)&=&k_B T^2\left( \frac{\partial \log Q}{\partial
T}\right)_{V,N},
\end{eqnarray}
where $Q=\sum_j \exp(-E_j/k_BT)$ is the canonical partition
function.

\begin{figure}[H]
\includegraphics[width=0.5\textwidth, clip=]{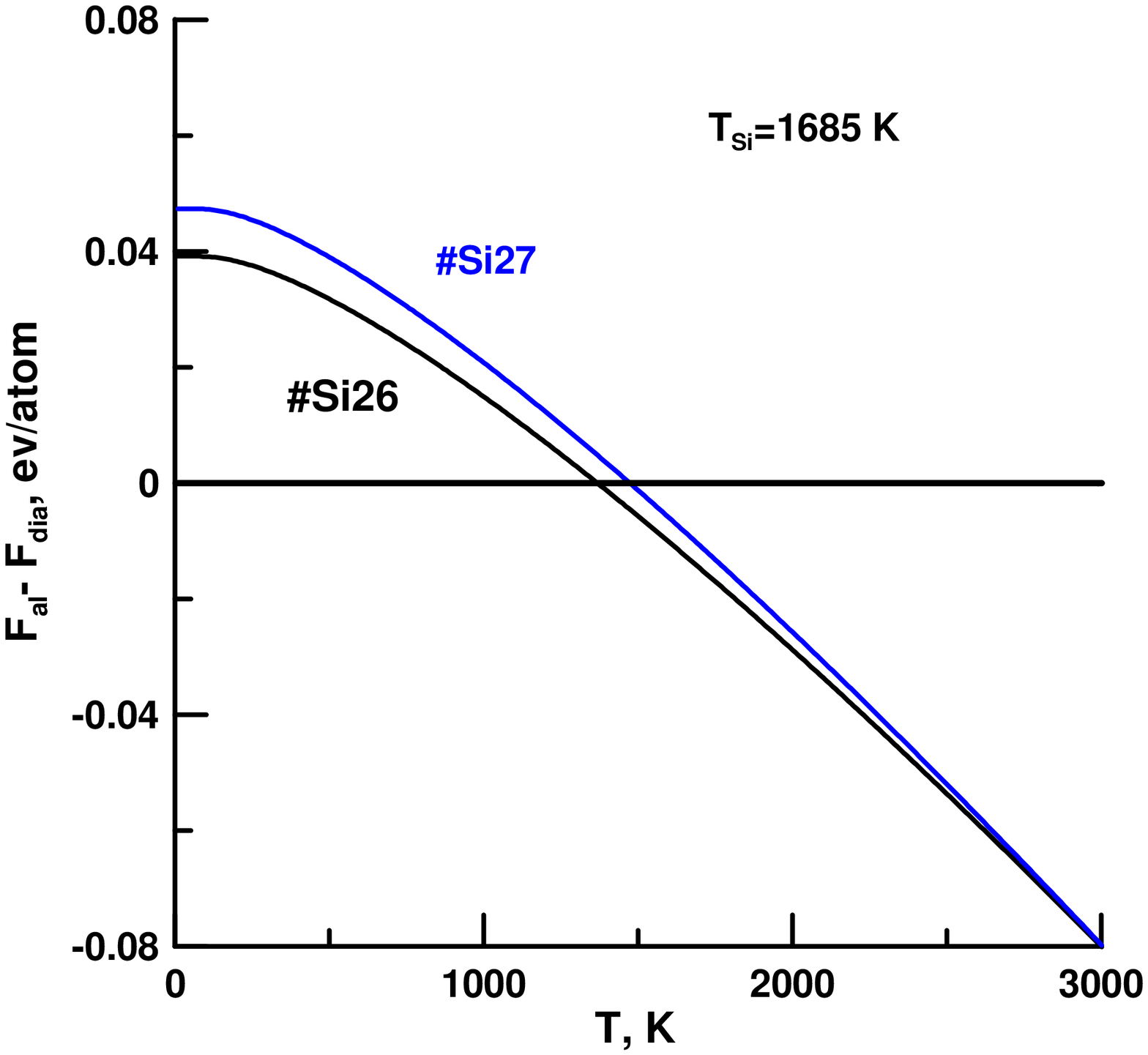}\includegraphics[width=0.5\textwidth, clip=]{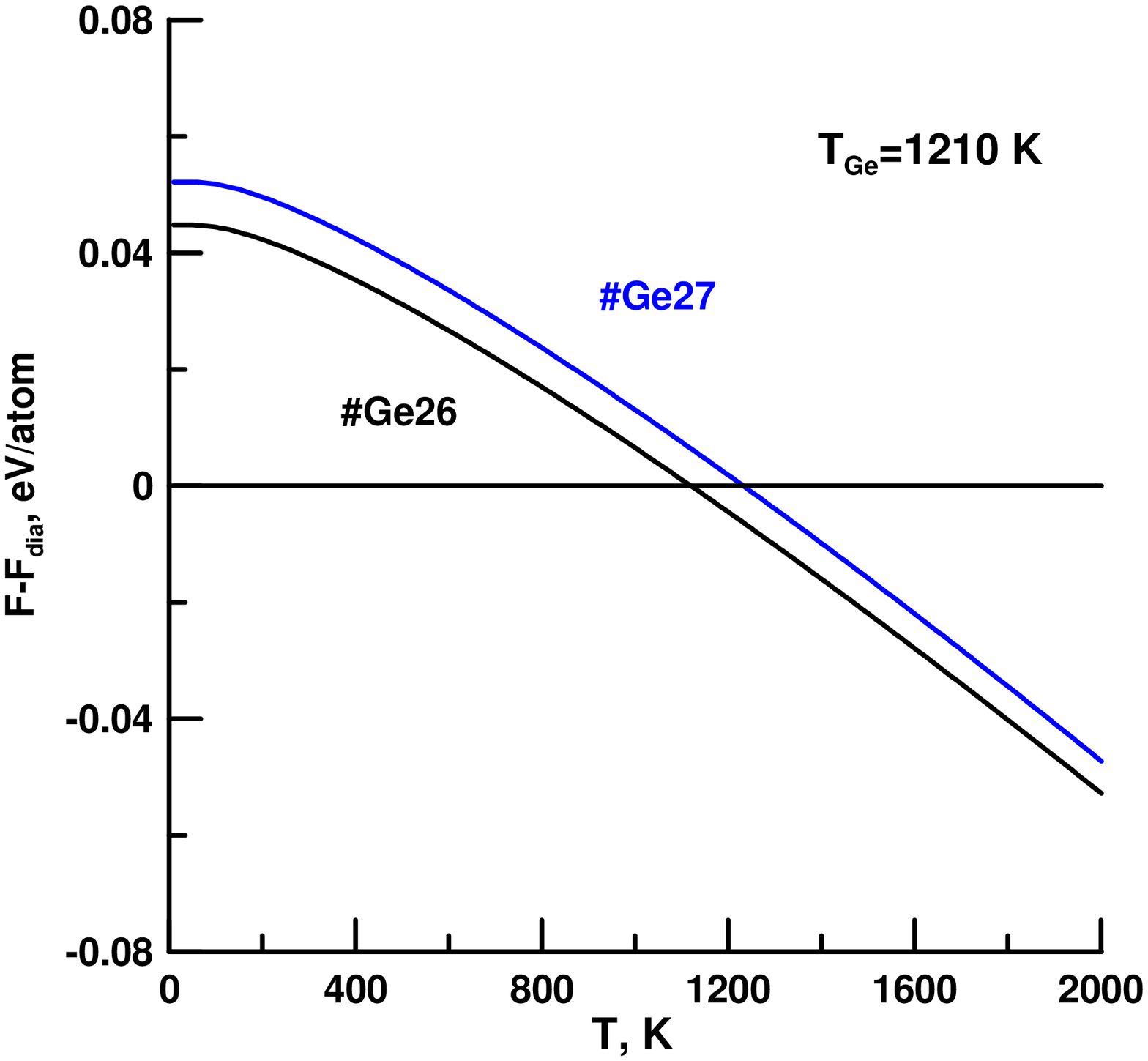}
\caption{Free energy difference $\Delta F(T)$ per atom as a
function of absolute temperature for silicon (left panel) and
germanium  (right panel) allotropes $\#26$ and
$\#27$.}\label{fig:FreSi}
\end{figure}

 We have performed these frequency calculations using a $2\times 2\times 2$
 supercell.
 We find that, contrary to the pressure dependence, the predicted Si and Ge
 allotropes may be high-temperature variants (see Fig. \ref{fig:FreSi}  for two
 examples of Si\#26, Si\#27 and Ge\#26, Ge\#27 allotropes), but because
 of the limitations of the quasi-harmonic method adopted here, it cannot be
 ascertained whether this is the case or not, as the temperatures of transition
 are quite close to the melting temperature of silicon and germanium (resp.
 \num{1687} and \SI{1211}{K} at $p=\SI{101325}{Pa}$,
see Ref.~\onlinecite{CRC96th}).

\subsection{Mechanical properties}
\label{sec:42} Silicon and germanium are not expected to exhibit
mechanical properties of interest; nevertheless, it is necessary to
compute the elastic constants $C_{ab}$ in order to check the
stability of the new allotropes against strain: all of them
were found to be stable according to the Cauchy-Born criterion, i.e.
the generic necessary and sufficient criterion that all eigenvalues
of matrix $C_{ab}$ be positive \cite{BornHuang1954,Cab}. The elastic constants
are presented in the Supplementary Material. The bulk moduli are remarkably
similar and systematically lower than the (already low) value for
the diamond structure. Exactly as for the diamond structure, Ge allotropes
exhibit lower bulk moduli~$B$ than their Si equivalents because of longer bonds,
implying a lower electronic bonding density~\cite{Gilman2003}. The present
allotropes exhibit a small
mechanical anisotropy, either in pure extension or in shear or both;
only structures $\#55$ and $\#88$ are slightly more compressible
along the [100] crystallographic axis, somewhat reminding
lonsdaleite (see below). The computed shear moduli $G$ are also lower
than for diamond: whereas for germanium the $B/G$ ratio remains close
to 1.35 (diamond Ge: 1.38), it is significantly higher for the Si
allotropes (1.7 against 1.47 for diamond Si), at the limit of the
Pugh threshold of 1.75 distinguishing brittle (low values of $B/G$)
from ductile materials \cite{Pugh} (see however Ref.~\onlinecite{QLong}). In
the same way, the Poisson ratio rises to 0.25-0.26 for the Si
allotropes (0.22 for diamond Si), which is generally interpreted as
another sign of increasing ductility \cite{QLong}, while remaining
around 0.2-0.21 for Ge (0.21 for diamond Ge): the variance between
Si and Ge is thus significantly larger for the new allotropes than
for diamond. The bulk and shear moduli for the predicted silicon and germanium
allotropes are accumulated in the Tables \ref{Table:2} and \ref{Table:3}. The
shear moduli were calculated via the elastic constants $C_{ab}$ using the Voigt
prescription \cite{Voigt}. The bulk moduli were calculated from the equations
of state as well as from the elastic constants, with very close resulting
values.\par
As Ge bonding electrons are more delocalized than their
Si counterparts in all the investigated allotropes (as can be seen for
instance in plots of electronic
charge densities or of the Electron Localization Function~\cite{ELF1,ELF2}, not shown here),
bonds in Ge are more susceptible to angular deformation and correspondingly the
shear moduli (measuring the resistance to shape changes) of the Ge compounds
are always smaller than those of their Si equivalents (see for instance
Ref.~\onlinecite{Gilman2003}). From that point of view, Ge is indeed
truly intermediate between C and Si (covalent bonding, strongly localized
electrons) and Sn and Pb (metallic bonding, highly delocalized electrons).
Consequently, the distortion to pure $sp^3$ bonding has more effects on~Si
allotropes.

\subsection{Vibrational properties}
\label{sec:43} Figs. \ref{fig:PDOSSi} and \ref{fig:PDOSGe} show (for
structure $\#28$; see Supplementary Material for the others) the
phonon spectra and Density of States (DOS) in the BZ: no
instabilities were found. As expected, the lowest dominant frequency
is higher for Si than for Ge, as the latter atom is heavier (72.63
against 28.08 g mol$^{-1}$); if provisions are made for the
different masses, the spectra are actually quite similar. The
predominance in the DOS of high-frequency modes indicates that
bond-bending modes are favored over bond-stretching; this is
confirmed by the value of the Kleinman parameter \cite{Kleiman}:
\begin{equation}
\zeta=\frac{C_{11}+8C_{12}}{7C_{11}+2C_{12}},
\end{equation}
which is low for all the allotropes (0 corresponding to
bond-bending, 1 to bond-stretching modes), especially for Ge (some
Si allotropes reach a value of $\frac{1}{2}$ and thus exhibit both
modes), see the discussion at the end of the previous paragraph. Notice also
that these high-frequency modes exhibit even lower dispersion across the BZ
than diamond.

\begin{figure}[H]
\includegraphics[width=0.5\textwidth, clip=]{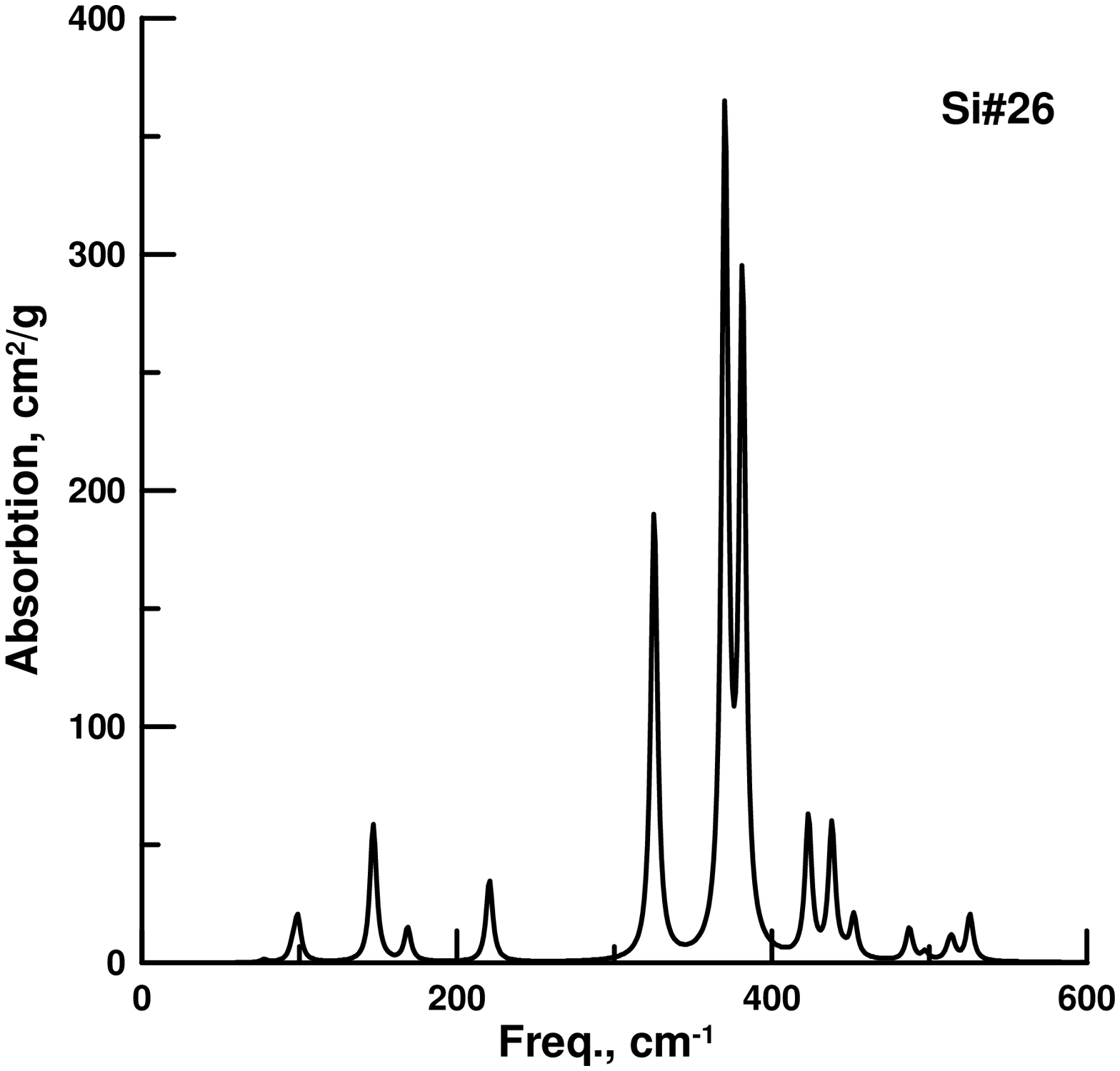}\includegraphics[width=0.5\textwidth, clip=]{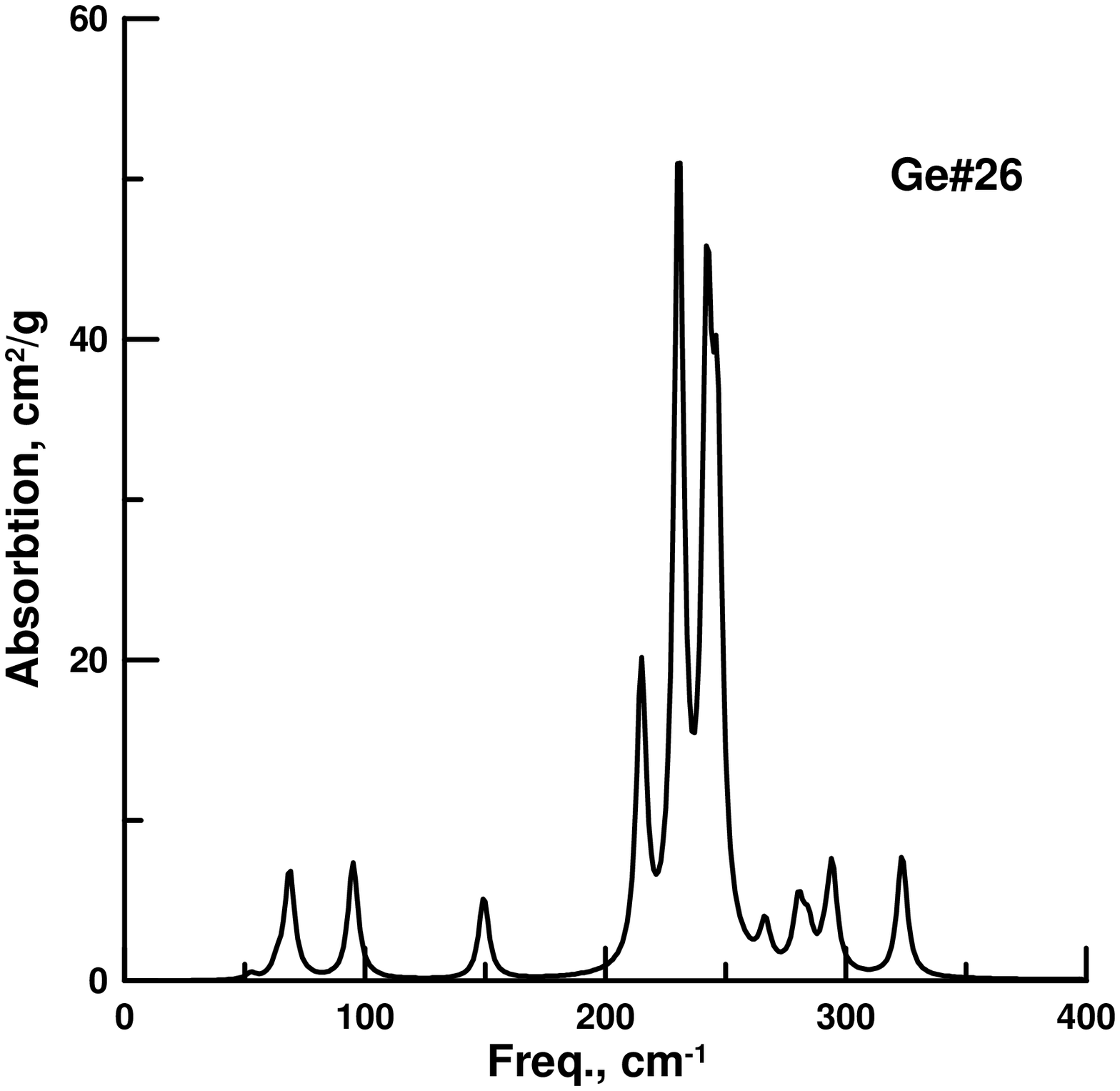}
\caption{IR absorbtion spectra for silicon (left panel) and
germanium (right panel) allotropes $\#26$ (black line)
.}\label{fig:IrSiGe}
\end{figure}

Closely connected to phonons are the Raman shift spectra (see Fig.
\ref{fig:IrSiGe}); the Si structure retains the complicated spectrum
of the C equivalent (see \onlinecite{BPSS}), whereas Ge has a far
simpler spectrum. In substance, as symmetry is the same in both
cases, the same vibrational modes are Raman-active (recall that
diamond has by symmetry only one Raman-active mode); however, for C
and Si allotropes, the derivative of the electronic polarisability
is clearly higher, i.e. the (static) electronic contribution to the
dielectric matrix is more susceptible to variations in atomic
positions; notice that defective diamond~\cite{Baima2016} shows a very similar
Raman spectrum. If for carbon this is a direct consequence of the very
short bonds (between 1.5 and 1.6 ${\AA}$), this mechanism cannot be
invoked for Si with respect to Ge, as bonds have basically the same
lengths (2.4-2.6 against 2.3-2.4 ${\AA}$ for Si). As a consequence,
the experimental characterization of the Ge allotrope should be easy
(a single shift), whereas it would be quite difficult for C or Si
allotropes.

Contrary to the ideal diamond structures of silicon and germanium,
the predicted allotropes absorb IR radiation, as shown in Fig.
\ref{fig:IrSiGe} for silicon and germanium allotropes $\#26$ (see
Supplementary Material for the others). The absorbance spectrum
$I(\nu)$ is calculated according to the following classical absorption formula,
averaged over the inequivalent polarization directions $ii$:
\begin{equation}
I(\nu)=\frac{1}{3}\sum_{ii=1}^3 \frac{4 \pi}{\lambda \rho} \mbox{Im
}[n_{ii}(\nu)]
\end{equation}
where $\lambda$ is the wavelength of the incident beam,  $\rho$
the crystal density of mass per unit volume and $n_{ii}(\nu)$ the complex
refractive index, which is obtained via the real and imaginary parts of the
complex dielectric tensor $\epsilon_{ii}(\nu)$, in turn computed for each
inequivalent polarization direction according to the classical
Drude-Lorentz model:
\begin{equation}
\epsilon_{ii}(\nu)=\epsilon_{opt}+\sum_p
\frac{f_{p,ii}\nu_p^2}{\nu_p^2-\nu^2-i \nu\gamma_p},
\end{equation}
where $\epsilon_{opt}$ is the optical (high-frequency) dielectric
tensor, $\nu_p$, $f_p$, $\gamma_p$ the frequency, oscillator
strength, and damping factor for the $p$-th vibration mode respectively.

\subsection{Electronic properties}
\label{sec:44} Figs. \ref{fig:BandSi} and \ref{fig:BandGe} show (for
structures $\#28$; see Supplementary Material for the others) the
electronic band structures and the DOS; the origin of energy is set
at the top of the valence band. There are again differences in
detail between Si and Ge, essentially confined to the conduction
band, just as for the diamond structure. The valence bands show on
average more similarities with lonsdaleite   than with
diamond\cite{De}: in particular the DOS is larger near the top of
the valence band with a distribution covering 4 eV, and the acute
maximum of the diamond structure (corresponding of course to strong $s$-$p$
overlap) is far less pronounced in all the
present allotropes. The same remark holds for the conduction band.
The width of the valence band is reduced with respect to the diamond
structure; this, and the deviation of the bonding angles with
respect to the ideal tetrahedral value (see \S~\ref{sec:46} below) are clear
markers of a weakened bonding; if the
contributions of the $s$ and $p$ electrons to the DOS are computed (not
shown here), we observe the quasi absence of overlap near the top of
the band (dominated by $p$ electrons), which is of course also seen in
the narrow and relatively flat profile of the top valence bands
(say, from -4 to 0 eV). This is expected as all the present
structures have higher values of energy than diamond: the deficit is
to be found in lower bonding energies. Of interest is the structure
\#27 of Ge, which exhibits a low indirect band gap of barely 0.23 eV
with the PBE exchange-correlation functional (see Table \ref{Table:3});
spin-orbit correction yields a difference of \SI{0.05}{eV}.
While this low value of the band gap is probably also due to the well-known
tendency of DFT-GGA to underestimate band gaps,
it is just as well the consequence of the weakening of the $s$-$p$
overlap, which suggests that some (yet unknown) phase might actually
be nearly or completely metallic, as it is the case for liquid
germanium~\cite{JLessCommonMet.145.531} and amorphous metallic
germanium~\cite{PhilosMag.29.547}. The band gaps obtained in
calculations with hybrid-type functional HSE06~\cite{HSE06} are also
presented in the Tables (\ref{Table:1})- (\ref{Table:3}). Let us
note that the presented values for direct ($\Delta E_{\Gamma}$) and
indirect band gaps were calculated using the full-electron basis set in
CRYSTAL14: the computations agree reasonably well with experimental data
for diamond structures in the PBE case, while they overestimate the same band
gaps in the HSE case. On the contrary, computations made with VASP show that HSE
computations agree well with the experimental data whereas PBE calculations
strongly underestimate the same data.~\cite{Hummer2009}

\subsection{Optical properties}
\label{sec:45} As mentioned in the Introduction, the absorption
and refraction spectra of the silicon and germanium allotropes
are of interest for photovoltaic applications. For this
purpose, we performed calculations of frequency-dependent dielectric
functions of the new proposed allotropes, both in standard DFT and
hybrid-functional frameworks. It is well-known that the former fails
to reproduce the optical band gaps, especially for low-gap
structures such as germanium, which is predicted to be a metal.
This closing of a band gap is due to the strong correlation of
$d$-electrons in germanium, which can be partly corrected by
implementing a hybrid functional with some part of exact
exchange. Hybrid functionals lead to reliable results for
the band gaps of a number of elemental and binary insulators and
semiconductors. In our calculations we have used the screened Heyd,
Scuseria, and Ernzerhof (HSE06)~\cite{HSE06} hybrid functional. We should
mention that we do not
pretend to describe the dielectric properties
 of the considered structures with high accuracy in this work,
but only to give a correct qualitative picture. Following this, we
will work in the random-phase approximation and neglect local fields
corrections.
To make accurate predictions one needs to take into account both
the local fields and excitonic effects; as is well-known, DFT is a theory
based on the ground state and therefore insufficient to describe the
excited states. For accurate absorption spectra with
excitonic effects one has to solve the Bethe-Salpeter equation
working in the GW approximation or in the framework of time-dependent DFT.

As implemented in the VASP package, the imaginary part of the
frequency-dependent dielectric tensor is defined by the following formula
\begin{equation}
\varepsilon_{\alpha\beta}^{(2)}(\omega)=\frac{4\pi^2
e^2}{\Omega}\lim_{q\to 0}\frac{1}{q^2}
\sum_{c,v,\mathbf{k}}2w_{\mathbf{k}}
\delta(\epsilon_{{c}\mathbf{k}}-\epsilon_{{v}\mathbf{k}}-\omega)\times\langle
u_{c\mathbf{k}+\mathbf{e}_\alpha q}| u_{v\mathbf{k}}\rangle\langle
u_{c\mathbf{k}+\mathbf{e}_\beta q}| u_{v\mathbf{k}}\rangle^*,
\end{equation}
where the indices $c$ and $v$ refer to conduction and valence band
states respectively in the sum over the empty states,
$\epsilon_{{v,c}\mathbf{k}}$ are the corresponding eigenenergies,
$\Omega$ the volume of a primitive cell, $k$-point weights
$w_{\mathbf{k}}$ are defined such that they sum to 1,
${e}_{\alpha(\beta)}$ are the unit vectors for the three Cartesian
directions, and  $ u_{c{\mathbf{k}}}$
 is the cell-periodic part of the orbitals at point $ \bf k$. The real
 part of the dielectric tensor  $ \varepsilon^{(1)}$ is obtained by the usual
 Kramers-Kronig transformation
\begin{equation}
\varepsilon_{\alpha\beta}^{(1)}(\omega)=1+\frac{2}\pi P
\int_{0}^\infty
\frac{\varepsilon_{\alpha\beta}^{(2)}(\omega\prime)\omega\prime}
{\omega\prime^2-\omega^2+i\eta}d\omega\prime
\end{equation}
where $P$ denotes the principal value. By cubic symmetry, the following holds
for diamond Si and Ge
\begin{equation}
\varepsilon_{xx}^{(1,2)}=\varepsilon_{yy}^{(1,2)}=\varepsilon_{zz}^{(1,2)},
\quad\varepsilon_{\alpha\beta}^{(1,2)}=0, \alpha\neq\beta,
\end{equation}
so the real and imaginary parts of the complex dielectric constant
$\varepsilon=\varepsilon_{1}+i\varepsilon_{2}$ can be determined by
$\varepsilon_{1,2}=\varepsilon_{xx}^{(1,2)}$. In the present case of
anisotropic structures, one can use the average values:
$\varepsilon_{1,2}=\bar\varepsilon_{1,2}=\dfrac{1}3(\varepsilon_{xx}^{(1,2)}+\varepsilon_{yy}^{(1,2)}+\varepsilon_{zz}^{(1,2)})$.

The real $(n)$ and imaginary $(k)$ parts of the complex refractive
index $n^*=n+ik$ define the refractive and absorption properties of a
crystal, and they can be determined using the following
relationships, as $\varepsilon=(n+ik)^2$:
\begin{eqnarray}
n&=&\left[\frac{(\varepsilon_1^2+\varepsilon_2^2)^{1/2}+\varepsilon_1}{2}\right]^{1/2} \\
k&=&\left[\frac{(\varepsilon_1^2+\varepsilon_2^2)^{1/2}-\varepsilon_1}{2}\right]^{1/2}.
\end{eqnarray}

To check the reliability of the obtained results we have also performed the
calculations for the diamond forms of Si and Ge, as experimental data for the
latter is known for a long time~\cite{SadaoAdachi}.
Cubic Si and Ge have a small, 2-atomic, primitive cell and a
correspondingly large BZ. To reduce the BZ we have
repeated the primitive cell twice along each independent direction,
introducing a supercell of $2\times 2\times 2$ containing $8\times 2=16$
atoms: the BZ is then reduced by a factor of 2 in all directions, allowing to
use half as many $k$-points in the calculations; this is crucial as
computations with hybrid functionals are very time-consuming compared to their
PBE equivalents, even more so as accurate results for optical properties
require a very dense $k$-point mesh of the order of 2 or 3 times that
typically used for relaxations. In the supercell approach just defined, a
$7\times7\times7$ Gamma-centered mesh is enough; moreover, this approach
eliminates the non-physical, artificial correlations 
arising from small fluctuations and errors in the electronic charge density on
the one hand and high symmetry and small primitive cells on the other hand: for
instance, by symmetry only 1/48 of the already small primitive cell of the
diamond structure is required.

\begin{figure}[H]
\includegraphics[width=0.5\textwidth, clip=]{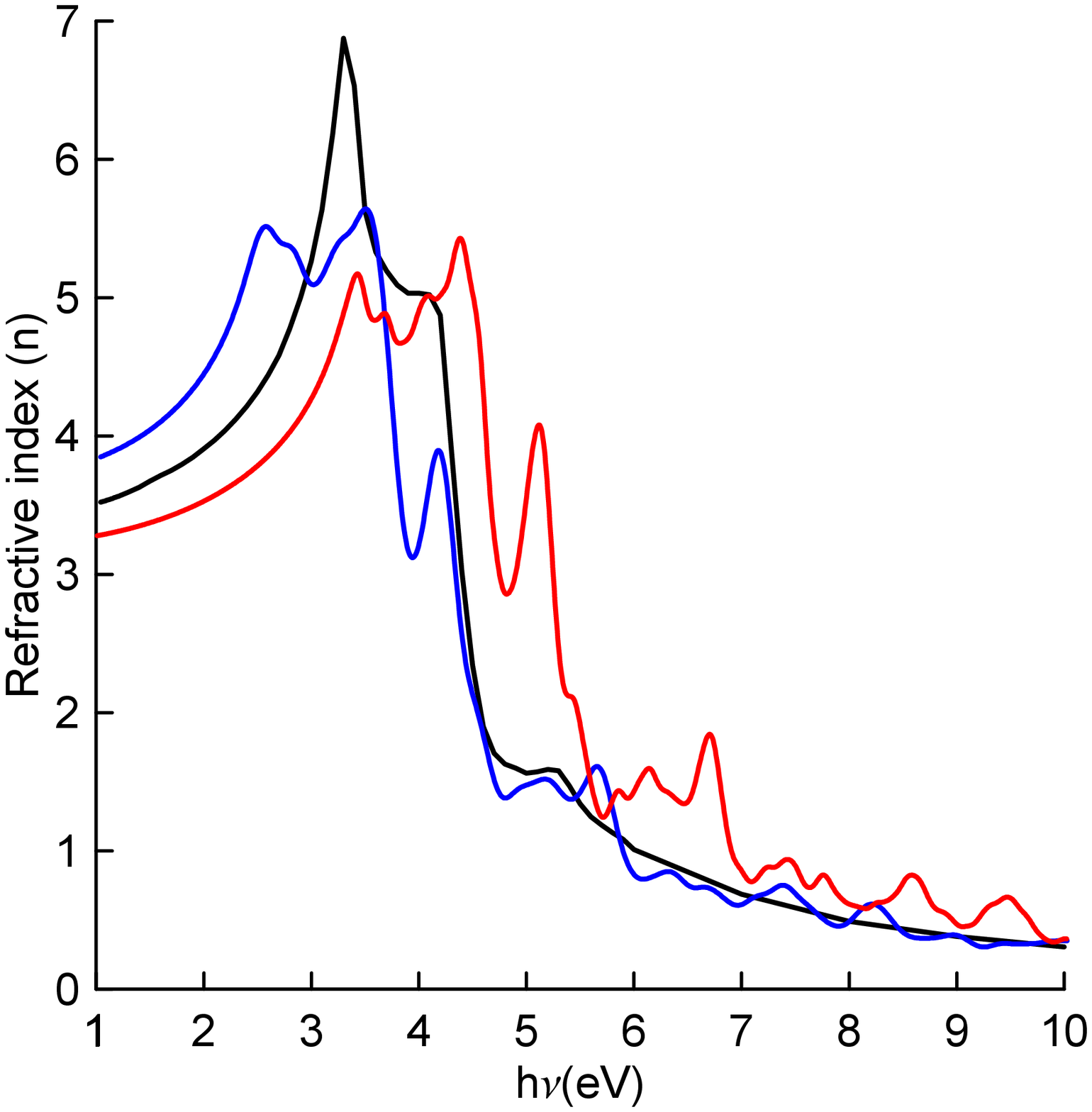}\includegraphics[width=0.5\textwidth, clip=]{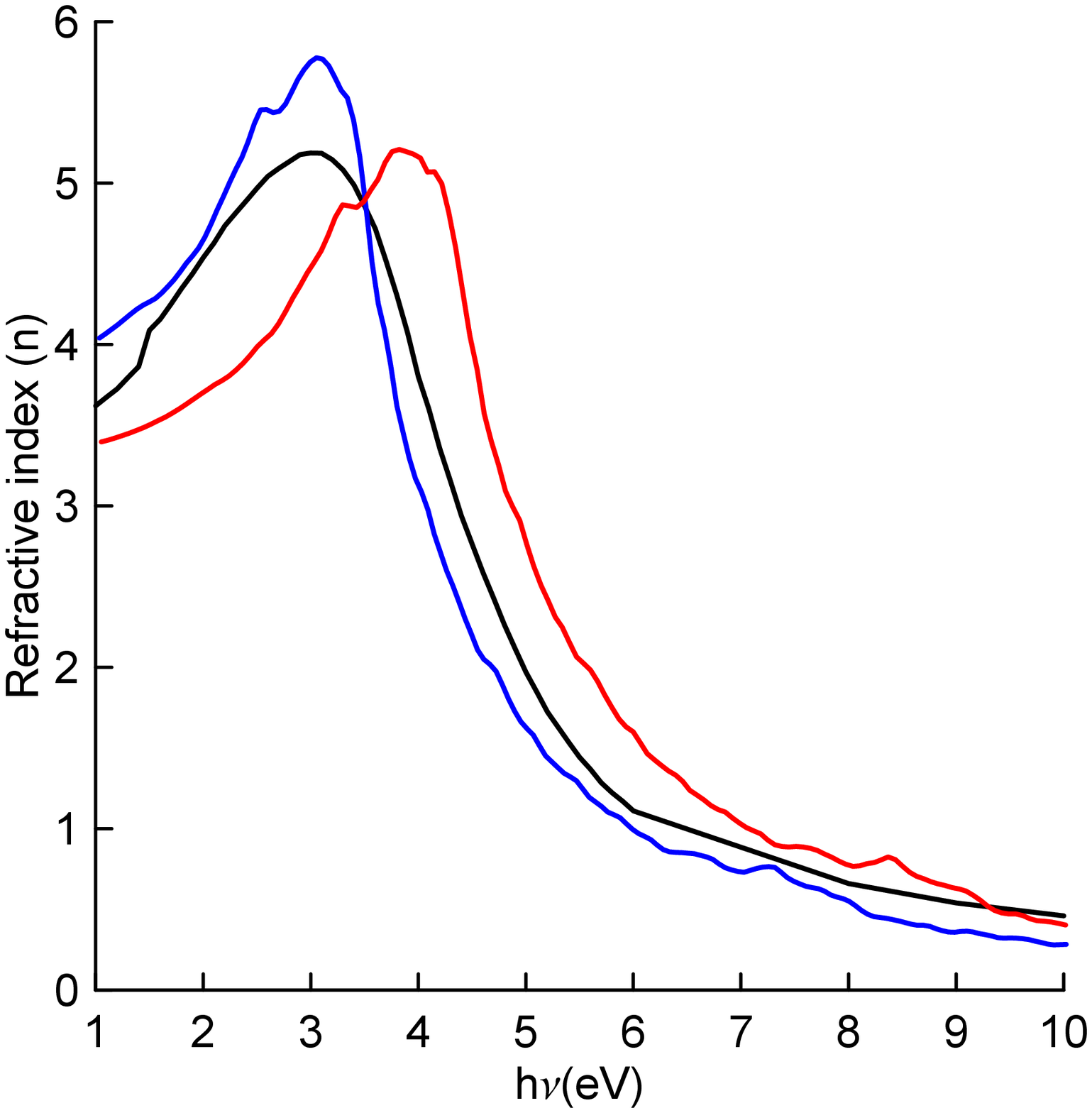}
\caption{Real part of the refractive index $(n)$ for Si diamond
(left panel) and Si$\#28$ (right panel) as a function of photon energy. Black
line -- experiment, blue line -- results obtained with PBE functional, red
line -- results obtained with HSE06 functional.}\label{fig:refrSi}
\end{figure}

\begin{figure}[H]
\includegraphics[width=0.5\textwidth, clip=]{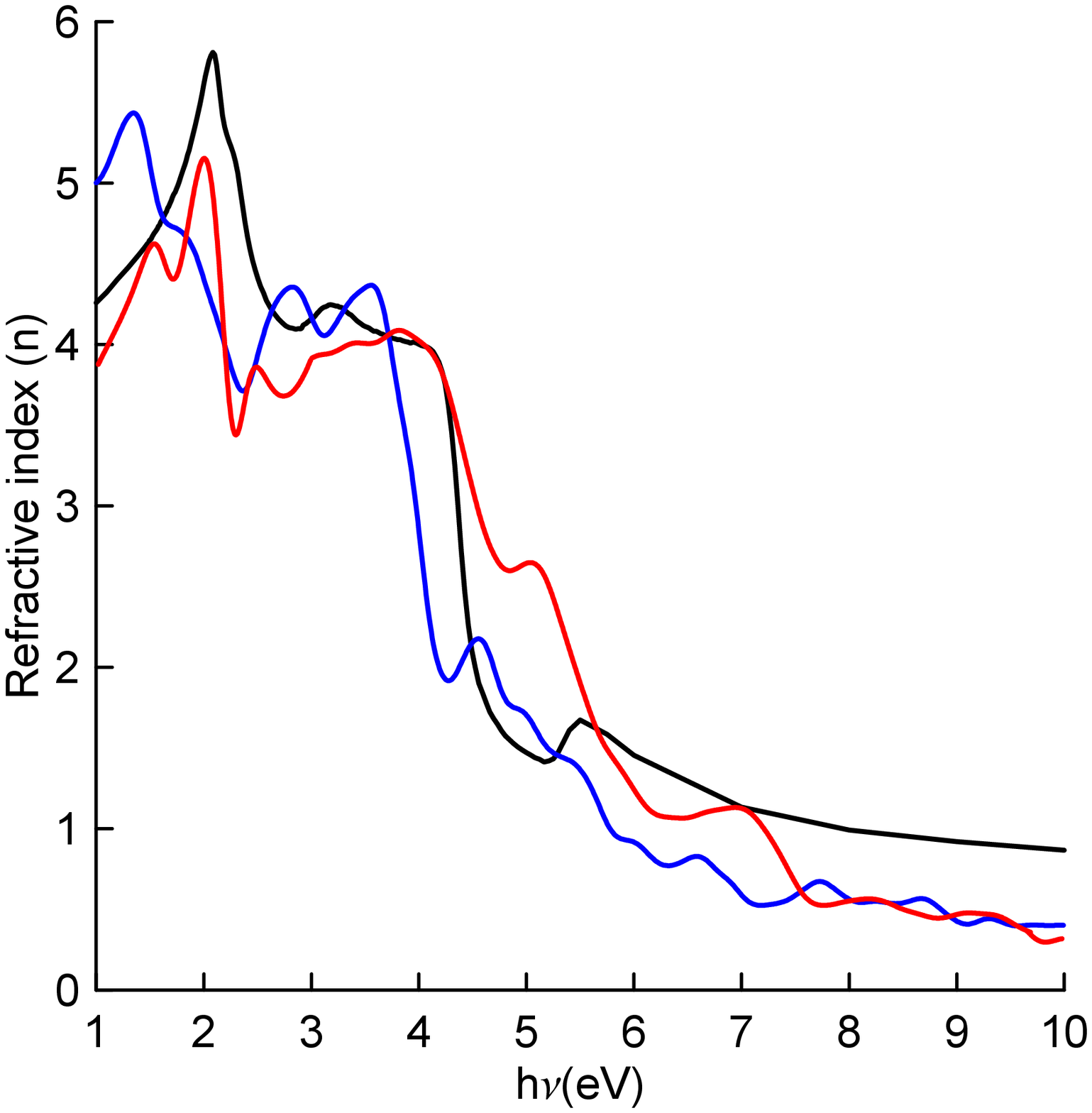}\includegraphics[width=0.5\textwidth, clip=]{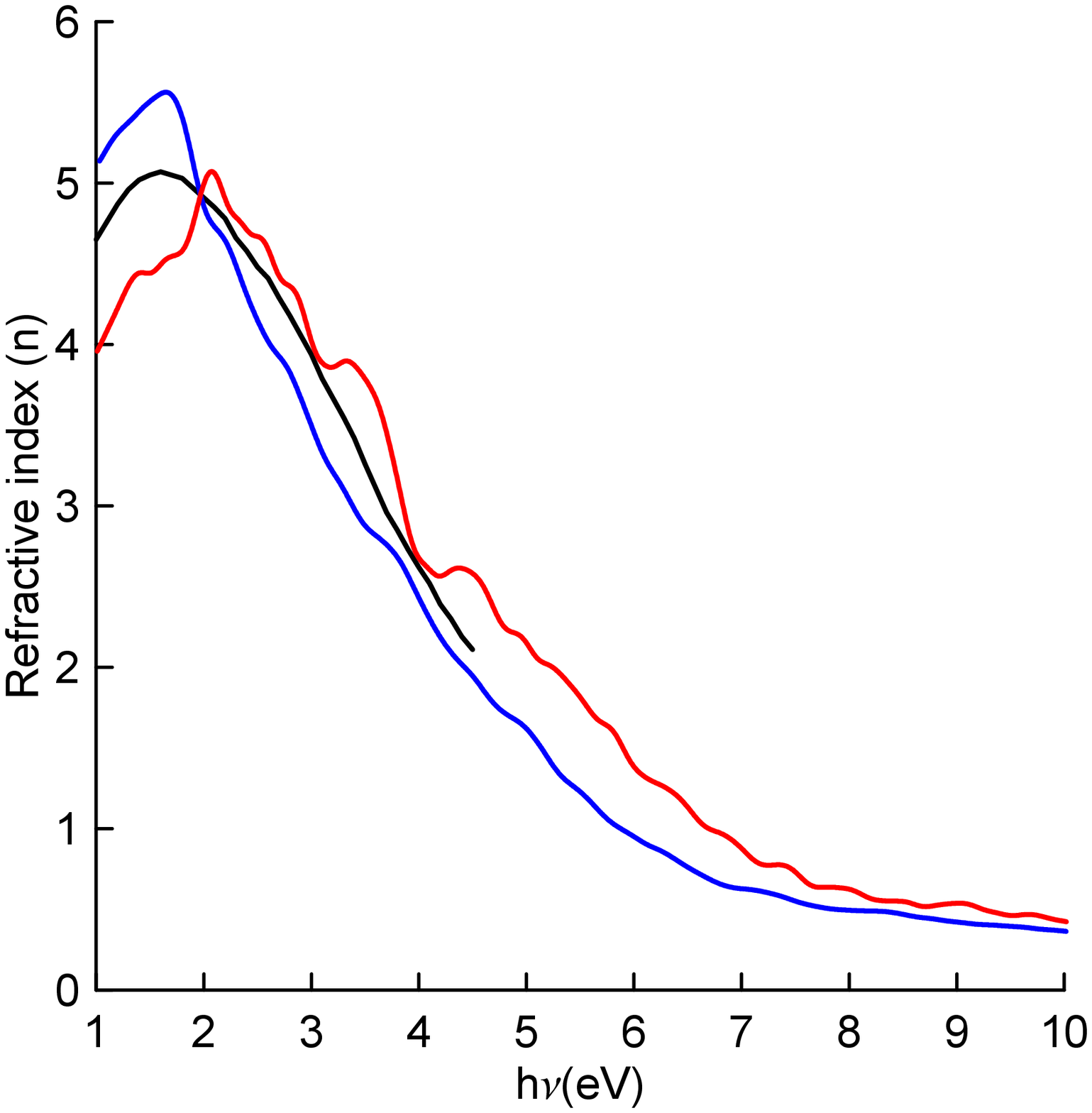}
\caption{Real part of the refractive index $(n)$ for Ge diamond (left panel)
and Ge$\#28$ (right panel) as a function of photon energy. Same legend as
in Fig.~\ref{fig:refrSi}.}\label{fig:refrGe}
\end{figure}

\begin{figure}[H]
\includegraphics[width=0.5\textwidth, clip=]{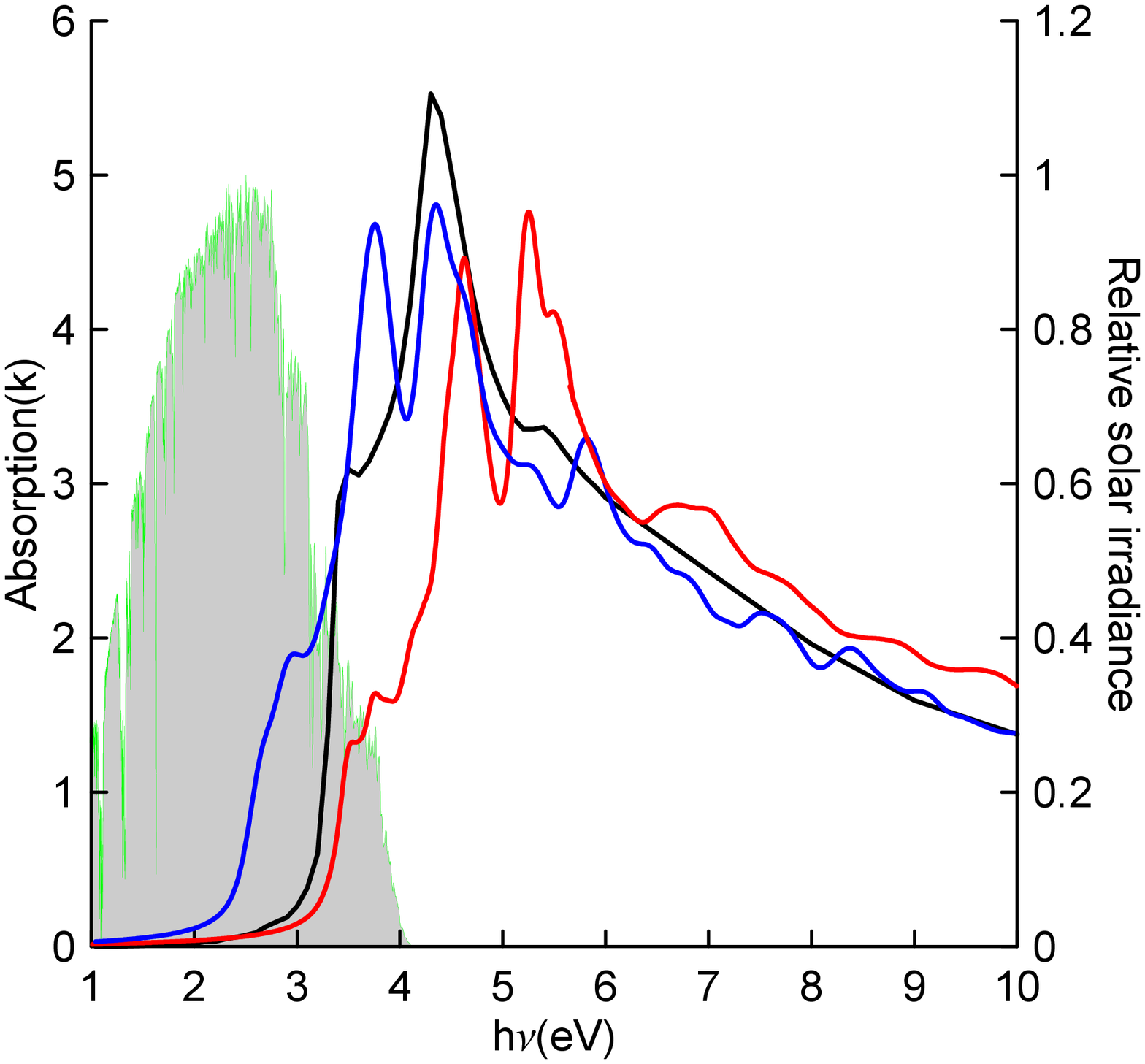}\includegraphics[width=0.5\textwidth, clip=]{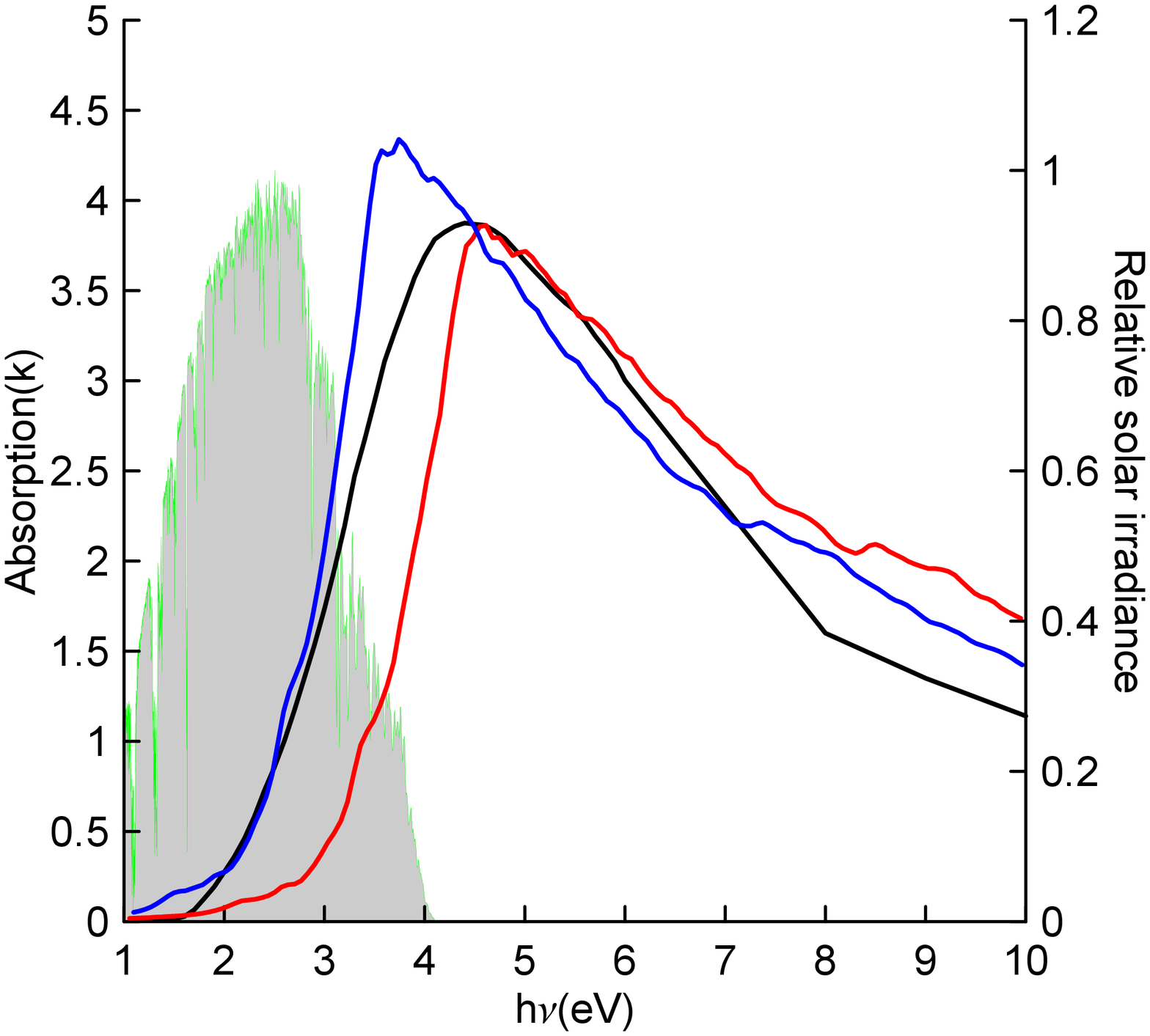}
\caption{Imaginary part of the refractive index $(k)$ for Si diamond (left
panel) and Si$\#28$ (right panel) as a function of photon energy. Black line --
experiment, blue line -- results obtained with PBE functional, red line --
results obtained with HSE06 functional. Green line with shaded area --
reference air mass 1.5 solar spectral irradiance.}\label{fig:absSi}
\end{figure}

\begin{figure}[H]
\includegraphics[width=0.5\textwidth, clip=]{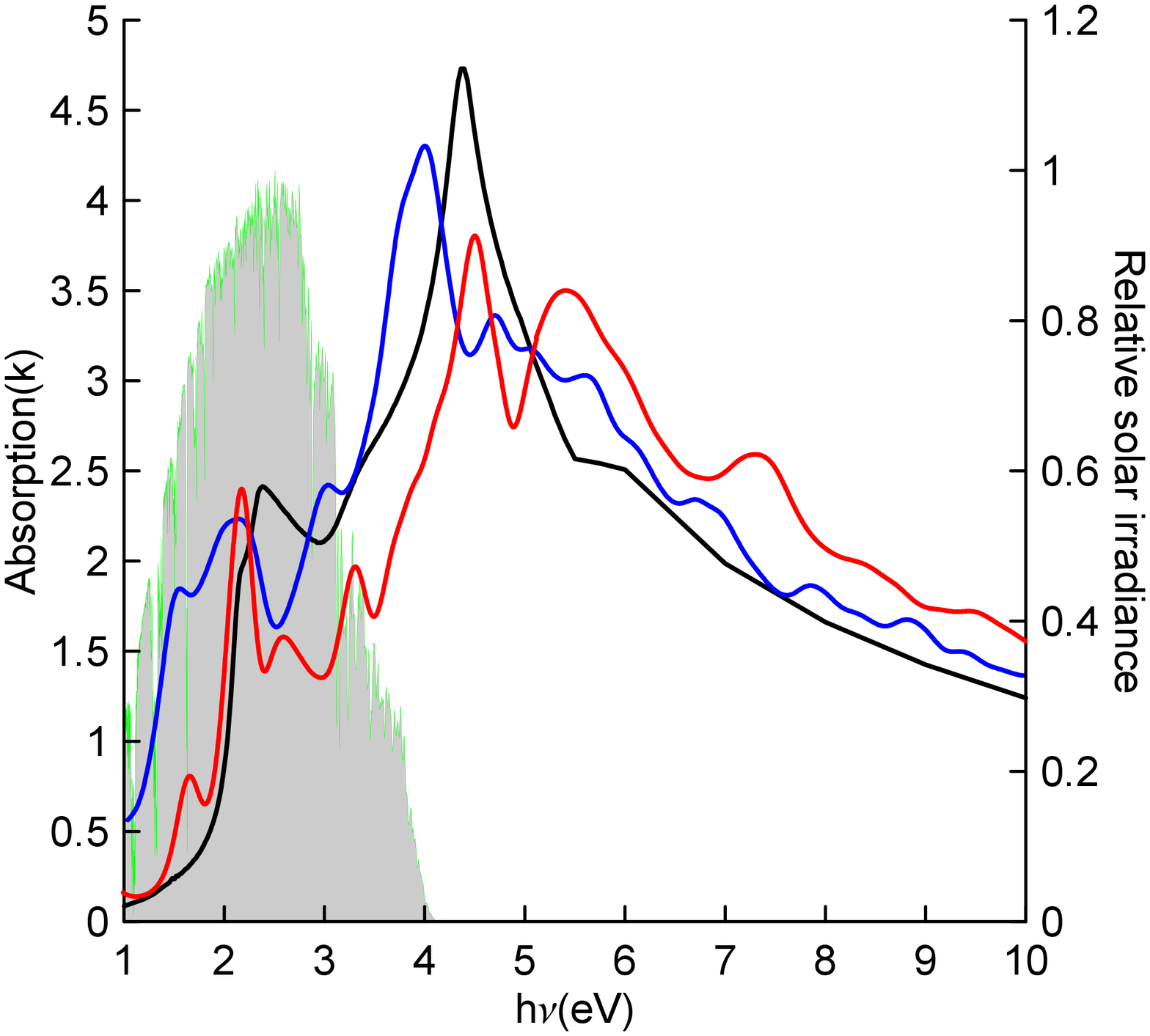}\includegraphics[width=0.5\textwidth, clip=]{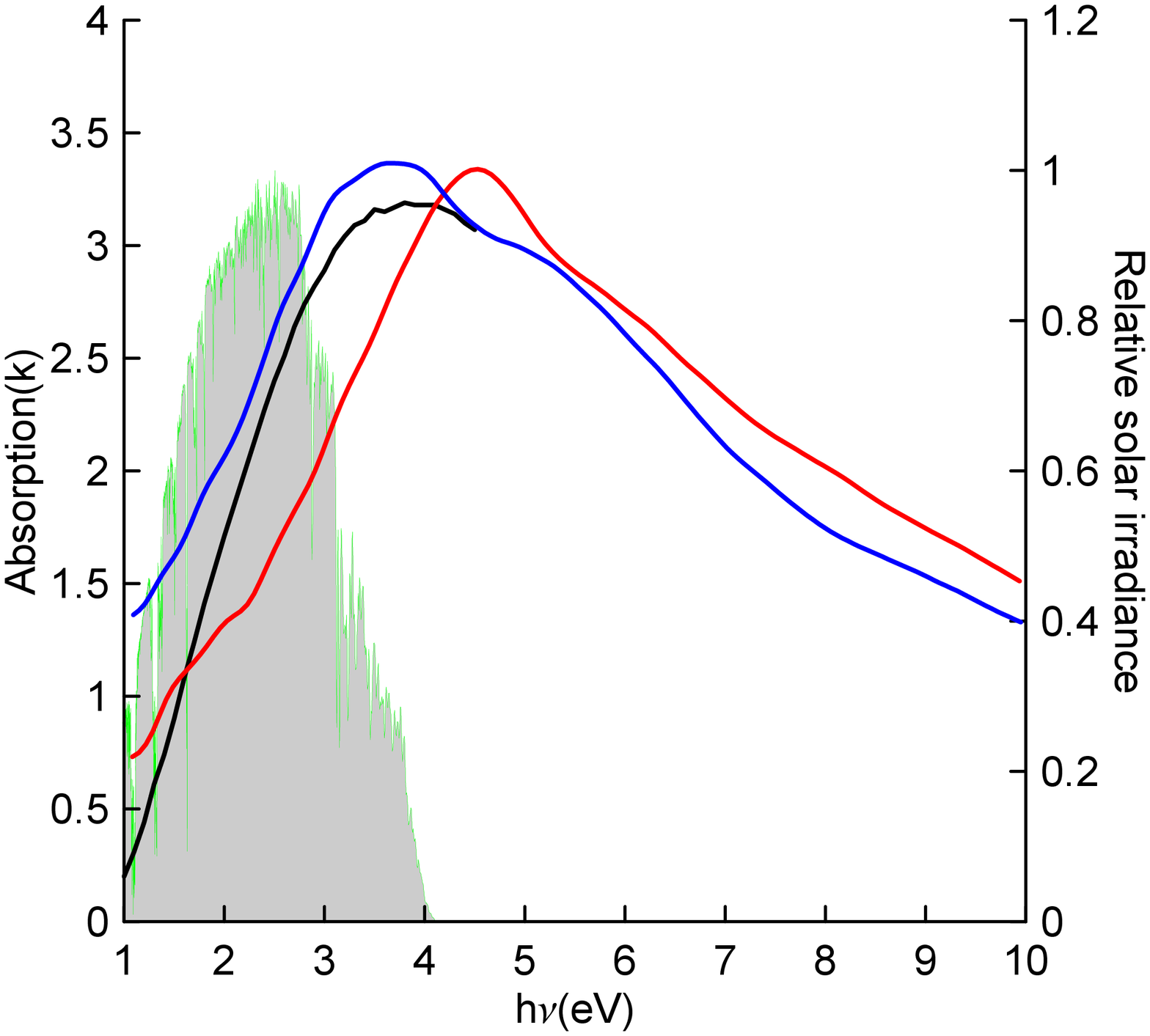}
\caption{Imaginary part of the refractive index $(k)$ for Ge diamond (left
panel) and Ge$\#28$ (right panel) as a function of photon energy. Same legend
as in Fig.~\ref{fig:absSi}.}\label{fig:absGe}
\end{figure}

In the left panels of Figs.~(\ref{fig:refrSi})
and~(\ref{fig:refrGe}) are shown the calculated (red line: HSE06, blue line:
PBE) and experimental~\cite{SadaoAdachi} (black line) frequency-dependent
refractive indices $n$ of diamond Si and Ge. The absorption spectra $(k)$
together with the reference air mass 1.5 solar spectral irradiance, given in
arbitrary units, are shown in the left panels of Figs.~(\ref{fig:absSi})
and~(\ref{fig:absGe}).
The HSE06 functional predicts the frequency dependence of $n$ and $k$
better than its PBE counterpart: refraction and absorption index peaks are
correctly positioned for silicon, whereas for germanium there is a tendency to
underestimate the experimental values; in general, we find an overall fair
agreement perturbed by some oscillations.\par
Amorphous forms of Si and Ge, the so-called $a-$Si and
$a-$Ge, their various hydrogenated forms and some Si-Ge alloys (see again
Ref.~\onlinecite{RenewSustEnergRev.15.2165}) have shown promising properties
for their use in electronics and photovoltaics, and in particular for solar
cells. An accurate investigation of optical functions of amorphous Si can be
found in Ref.~\onlinecite{FerlautoAmorph}; their optical spectra are
characterized by one smooth absorption peak.
If we compare the calculated refractive index and absorption spectra of our
predicted allotropes with the corresponding spectra of amorphous forms (see the
right panels of Figs.~\ref{fig:refrSi}, \ref{fig:absSi} and
Figs.~\ref{fig:refrGe}, \ref{fig:absGe} respectively), we find a quantitative
and qualitative agreement with the experimental data both in the positions and
in the numerical values of the refractive and absorption peaks. This is strong
evidence that our predicted allotropes can be used as a counterpart of these
amorphous forms; indeed, allotrope Si $\#28$ demonstrates prominent absorbance
properties of solar light, and may thus be a promising crystalline structure
for thin-film solar cells.


\subsection{Structure and physical properties}
\label{sec:46} Looking at Table \ref{Table:4}, we notice two main
departures from the diamond structure: first, bond lengths are
distributed around their average value in diamond (2.35 ${\AA}$ in
Si, 2.45 ${\AA}$ in Ge). Second, bond angles are widely distributed
around the ideal tetrahedral value of 109.47$^\circ$, evocative of
distorted (and thus weakened) bonding; it is as if the
high-frequency phonon modes (of the bond-bending type) were actually
"frozen" in the structures. This should not be surprising, given the
proximity to the diamond and lonsdaleite phases, and the fact that
small "topological flips" might transform the latter into the
present allotropes (for instance two adjacent 6-rings may become
adjacent 5- and 7-rings), see Ref.~\onlinecite{BPSS}. The tiling approach
described there can also be used, and (given the structural
identities) leads to the same conclusions, namely that the
structures are either columns of adamantane cages ($\#28$ and
$\#50$), a monolayer of adamantane cages interconnected by sheets of
corrugated graphene ($\#88$), or built of lonsdaleite-like cages
(also interconnected by sheets of corrugated graphene in the case of
the two monoclinic allotropes $\#26$ and $\#27$). In other words,
structures $\#28$, $\#50$, and $\#88$ share more similarities with
diamond than the other three, which can be seen in their
systematically larger band gaps. That being said, the electronic DOS is more
similar to lonsdaleite than diamond \cite{De} for all the
allotropes, as we have seen above, showing the limitations of
structural/topological analysis alone. Exactly as for the equivalent
carbon allotropes, the structures can be seen as continuous random
networks, explaining why some optical properties look very much like
amorphous silicon and germanium, see again
\onlinecite{JLessCommonMet.145.531,PhilosMag.29.547} for the latter case.

\section{Summary and conclusions}

Six new allotropes of silicon and germanium, isostructural with the six carbon
allotropes studied in \onlinecite{BPSS}, have been investigated and their main
physical properties determined by first-principles computations. While
superficially similar, resulting mainly of their close proximity to their
diamond/lonsdaleite parent structure, these silicon and germanium allotropes
show actually some differences in detail. Severe structural distortions
have as main consequence the weakening of the $s$-$p$ overlap characteristic of
the pure covalent bonding present in the diamond structures. Along with the
structural similarity with the amorphous phases of Si and Ge and the known
metallic properties of Ge~\cite{JLessCommonMet.145.531,PhilosMag.29.547}, this
may suggest that some allotropes of Ge might also be metallic (even though none
of those investigated here is). Finally, as seen in Figs.~\ref{fig:HSi}
and~\ref{fig:HGe}, these phases might actually be high-temperature variants,
quite in agreement with the metallic behavior of liquid germanium.

\section*{Acknowledgements}

The work of V.A.S., A.V.S., and D.M.P. was financially supported by
the Megagrant number 14.25.31.0005 of the Russian Ministry of
Education and Science. The work of V.A.S. and A.V.S. was also supported
by the Ministry of Education and Science of Russia under
Competitiveness Enhancement Program of the Samara National Research
University for 2013-2020. The work of G.F. was supported by the
Samara National Research University in the framework of Task
05B-P010-073.

\label{sec:5}
\renewcommand{\theenumi}{(\Asbuk{enumi})}
\section*{Appendix A: The structure of supplementary material }

The electronic supplementary material to the paper consists of
following subsections

\begin{enumerate}[label=\Alph*.]
\item Crystallographic data for the silicon allotropes
\item Crystallographic data for the germanium allotropes
\item Matrices of the elastic constants $C_{ab}$ for silicon and germanium allotropes
\item Phonon band structure for silicon and germanium allotropes
\item Raman shift spectra for silicon and germanium allotropes
\item IR absorbtion spectra for silicon and germanium allotropes
\item Electronic bands and DOS for silicon and germanium allotropes
\end{enumerate}

\bibliographystyle{apsrev4-1}
%
\bibliography{Paper_arxive}

\begin{thebibliography}{48}%
\makeatletter
\providecommand \@ifxundefined [1]{%
 \@ifx{#1\undefined}
}%
\providecommand \@ifnum [1]{%
 \ifnum #1\expandafter \@firstoftwo
 \else \expandafter \@secondoftwo
 \fi
}%
\providecommand \@ifx [1]{%
 \ifx #1\expandafter \@firstoftwo
 \else \expandafter \@secondoftwo
 \fi
}%
\providecommand \natexlab [1]{#1}%
\providecommand \enquote  [1]{``#1''}%
\providecommand \bibnamefont  [1]{#1}%
\providecommand \bibfnamefont [1]{#1}%
\providecommand \citenamefont [1]{#1}%
\providecommand \href@noop [0]{\@secondoftwo}%
\providecommand \href [0]{\begingroup \@sanitize@url \@href}%
\providecommand \@href[1]{\@@startlink{#1}\@@href}%
\providecommand \@@href[1]{\endgroup#1\@@endlink}%
\providecommand \@sanitize@url [0]{\catcode `\\12\catcode `\$12\catcode
  `\&12\catcode `\#12\catcode `\^12\catcode `\_12\catcode `\%12\relax}%
\providecommand \@@startlink[1]{}%
\providecommand \@@endlink[0]{}%
\providecommand \url  [0]{\begingroup\@sanitize@url \@url }%
\providecommand \@url [1]{\endgroup\@href {#1}{\urlprefix }}%
\providecommand \urlprefix  [0]{URL }%
\providecommand \Eprint [0]{\href }%
\providecommand \doibase [0]{http://dx.doi.org/}%
\providecommand \selectlanguage [0]{\@gobble}%
\providecommand \bibinfo  [0]{\@secondoftwo}%
\providecommand \bibfield  [0]{\@secondoftwo}%
\providecommand \translation [1]{[#1]}%
\providecommand \BibitemOpen [0]{}%
\providecommand \bibitemStop [0]{}%
\providecommand \bibitemNoStop [0]{.\EOS\space}%
\providecommand \EOS [0]{\spacefactor3000\relax}%
\providecommand \BibitemShut  [1]{\csname bibitem#1\endcsname}%
\let\auto@bib@innerbib\@empty
\bibitem [{\citenamefont {Baburin}\ \emph {et~al.}(2015)\citenamefont
  {Baburin}, \citenamefont {Proserpio}, \citenamefont {Saleev},\ and\
  \citenamefont {Shipilova}}]{BPSS}%
  \BibitemOpen
  \bibfield  {author} {\bibinfo {author} {\bibfnamefont {I.~A.}\ \bibnamefont
  {Baburin}}, \bibinfo {author} {\bibfnamefont {D.~M.}\ \bibnamefont
  {Proserpio}}, \bibinfo {author} {\bibfnamefont {V.~A.}\ \bibnamefont
  {Saleev}}, \ and\ \bibinfo {author} {\bibfnamefont {A.}~\bibnamefont
  {Shipilova}},\ }\href@noop {} {\bibfield  {journal} {\bibinfo  {journal}
  {Phys. Chem. Chem. Phys.}\ }\textbf {\bibinfo {volume} {17}},\ \bibinfo
  {pages} {1332} (\bibinfo {year} {2015})}\BibitemShut {NoStop}%
\bibitem [{\citenamefont {Oganov}\ and\ \citenamefont
  {Glass}(2006)}]{JChemPhys.124.244704}%
  \BibitemOpen
  \bibfield  {author} {\bibinfo {author} {\bibfnamefont {A.~R.}\ \bibnamefont
  {Oganov}}\ and\ \bibinfo {author} {\bibfnamefont {C.~W.}\ \bibnamefont
  {Glass}},\ }\href@noop {} {\bibfield  {journal} {\bibinfo  {journal} {J.
  Chem. Phys.}\ }\textbf {\bibinfo {volume} {124}},\ \bibinfo {pages} {244704}
  (\bibinfo {year} {2006})}\BibitemShut {NoStop}%
\bibitem [{\citenamefont {Oganov}(2011)}]{Oganov2011}%
  \BibitemOpen
  \bibinfo {editor} {\bibfnamefont {A.~R.}\ \bibnamefont {Oganov}},\ ed.,\
  \href@noop {} {\emph {\bibinfo {title} {Modern {M}ethods of {C}rystal
  {S}tructure {P}rediction}}}\ (\bibinfo  {publisher} {Wiley-VCH},\ \bibinfo
  {address} {Berlin},\ \bibinfo {year} {2011})\BibitemShut {NoStop}%
\bibitem [{\citenamefont {Blatov}\ and\ \citenamefont
  {Proserpio}(2011)}]{BPinOganov2011}%
  \BibitemOpen
  \bibfield  {author} {\bibinfo {author} {\bibfnamefont {V.~A.}\ \bibnamefont
  {Blatov}}\ and\ \bibinfo {author} {\bibfnamefont {D.~M.}\ \bibnamefont
  {Proserpio}},\ }in\ \href@noop {} {\emph {\bibinfo {booktitle} {Modern
  {M}ethods of {C}rystal {S}tructure {P}rediction}}},\ \bibinfo {editor}
  {edited by\ \bibinfo {editor} {\bibfnamefont {A.~R.}\ \bibnamefont
  {Oganov}}}\ (\bibinfo  {publisher} {Wiley-VCH},\ \bibinfo {address}
  {Berlin},\ \bibinfo {year} {2011})\ pp.\ \bibinfo {pages} {1--28}\BibitemShut
  {NoStop}%
\bibitem [{\citenamefont {Blatov}\ \emph {et~al.}(2014)\citenamefont {Blatov},
  \citenamefont {Shevchenko},\ and\ \citenamefont
  {Proserpio}}]{CrystGrowthDes.14.3576}%
  \BibitemOpen
  \bibfield  {author} {\bibinfo {author} {\bibfnamefont {V.~A.}\ \bibnamefont
  {Blatov}}, \bibinfo {author} {\bibfnamefont {A.~P.}\ \bibnamefont
  {Shevchenko}}, \ and\ \bibinfo {author} {\bibfnamefont {D.~M.}\ \bibnamefont
  {Proserpio}},\ }\href@noop {} {\bibfield  {journal} {\bibinfo  {journal}
  {Cryst. Growth Des.}\ }\textbf {\bibinfo {volume} {14}},\ \bibinfo {pages}
  {3576} (\bibinfo {year} {2014})}\BibitemShut {NoStop}%
\bibitem [{\citenamefont {Deem}\ \emph {et~al.}(2009)\citenamefont {Deem},
  \citenamefont {Pophale}, \citenamefont {Cheeseman},\ and\ \citenamefont
  {Earl}}]{JPhysChem.113.21353}%
  \BibitemOpen
  \bibfield  {author} {\bibinfo {author} {\bibfnamefont {M.~W.}\ \bibnamefont
  {Deem}}, \bibinfo {author} {\bibfnamefont {R.}~\bibnamefont {Pophale}},
  \bibinfo {author} {\bibfnamefont {P.~A.}\ \bibnamefont {Cheeseman}}, \ and\
  \bibinfo {author} {\bibfnamefont {D.~J.}\ \bibnamefont {Earl}},\ }\href@noop
  {} {\bibfield  {journal} {\bibinfo  {journal} {J. Phys. Chem.}\ }\textbf
  {\bibinfo {volume} {113}},\ \bibinfo {pages} {21353} (\bibinfo {year}
  {2009})},\ \bibinfo {note} {see
  \url{http://www.hypotheticalzeolites.net/DATABASE/DEEM/DEEM_PCOD/index.php}}\BibitemShut
  {NoStop}%
\bibitem [{\citenamefont {Treacy}\ \emph {et~al.}(1997)\citenamefont {Treacy},
  \citenamefont {Randall}, \citenamefont {Rao}, \citenamefont {Perry},\ and\
  \citenamefont {Chadi}}]{ZKristallogr.212.768}%
  \BibitemOpen
  \bibfield  {author} {\bibinfo {author} {\bibfnamefont {M.~M.~J.}\
  \bibnamefont {Treacy}}, \bibinfo {author} {\bibfnamefont {K.~H.}\
  \bibnamefont {Randall}}, \bibinfo {author} {\bibfnamefont {S.}~\bibnamefont
  {Rao}}, \bibinfo {author} {\bibfnamefont {J.~A.}\ \bibnamefont {Perry}}, \
  and\ \bibinfo {author} {\bibfnamefont {D.~J.}\ \bibnamefont {Chadi}},\
  }\href@noop {} {\bibfield  {journal} {\bibinfo  {journal} {Z. Kristallogr.}\
  }\textbf {\bibinfo {volume} {212}},\ \bibinfo {pages} {768} (\bibinfo {year}
  {1997})},\ \bibinfo {note} {see
  \url{http://www.hypotheticalzeolites.net/DATABASE/BRONZE_CONFIRMED/index.html}}\BibitemShut
  {NoStop}%
\bibitem [{\citenamefont {Bhubaneswari}\ \emph {et~al.}(2011)\citenamefont
  {Bhubaneswari}, \citenamefont {Iniyan},\ and\ \citenamefont
  {Goic}}]{RenewSustEnergRev.15.1625}%
  \BibitemOpen
  \bibfield  {author} {\bibinfo {author} {\bibfnamefont {P.}~\bibnamefont
  {Bhubaneswari}}, \bibinfo {author} {\bibfnamefont {S.}~\bibnamefont
  {Iniyan}}, \ and\ \bibinfo {author} {\bibfnamefont {R.}~\bibnamefont
  {Goic}},\ }\href@noop {} {\bibfield  {journal} {\bibinfo  {journal} {Renew.
  Sust. Energ. Rev.}\ }\textbf {\bibinfo {volume} {15}},\ \bibinfo {pages}
  {1625} (\bibinfo {year} {2011})}\BibitemShut {NoStop}%
\bibitem [{\citenamefont {El~Chaar}\ \emph {et~al.}(2011)\citenamefont
  {El~Chaar}, \citenamefont {Lamont},\ and\ \citenamefont
  {El~Zein}}]{RenewSustEnergRev.15.2165}%
  \BibitemOpen
  \bibfield  {author} {\bibinfo {author} {\bibfnamefont {L.}~\bibnamefont
  {El~Chaar}}, \bibinfo {author} {\bibfnamefont {L.~A.}\ \bibnamefont
  {Lamont}}, \ and\ \bibinfo {author} {\bibfnamefont {N.}~\bibnamefont
  {El~Zein}},\ }\href@noop {} {\bibfield  {journal} {\bibinfo  {journal}
  {Renew. Sust. Energ. Rev.}\ }\textbf {\bibinfo {volume} {15}},\ \bibinfo
  {pages} {2165} (\bibinfo {year} {2011})}\BibitemShut {NoStop}%
\bibitem [{\citenamefont {Zwijnenburg}\ \emph {et~al.}(2010)\citenamefont
  {Zwijnenburg}, \citenamefont {Jelfs},\ and\ \citenamefont
  {Bromley}}]{Zwijnenberg2010}%
  \BibitemOpen
  \bibfield  {author} {\bibinfo {author} {\bibfnamefont {M.}~\bibnamefont
  {Zwijnenburg}}, \bibinfo {author} {\bibfnamefont {K.}~\bibnamefont {Jelfs}},
  \ and\ \bibinfo {author} {\bibfnamefont {S.}~\bibnamefont {Bromley}},\
  }\href@noop {} {\bibfield  {journal} {\bibinfo  {journal} {Phys. Chem. Chem.
  Phys.}\ }\textbf {\bibinfo {volume} {12}},\ \bibinfo {pages} {8505} (\bibinfo
  {year} {2010})}\BibitemShut {NoStop}%
\bibitem [{\citenamefont {Selli}\ \emph {et~al.}(2012)\citenamefont {Selli},
  \citenamefont {Baburin}, \citenamefont {Martanak},\ and\ \citenamefont
  {Leoni}}]{Baburin2012}%
  \BibitemOpen
  \bibfield  {author} {\bibinfo {author} {\bibfnamefont {D.}~\bibnamefont
  {Selli}}, \bibinfo {author} {\bibfnamefont {I.}~\bibnamefont {Baburin}},
  \bibinfo {author} {\bibfnamefont {R.}~\bibnamefont {Martanak}}, \ and\
  \bibinfo {author} {\bibfnamefont {S.}~\bibnamefont {Leoni}},\ }\href@noop {}
  {\bibfield  {journal} {\bibinfo  {journal} {Sci. Rep.}\ }\textbf {\bibinfo
  {volume} {3}},\ \bibinfo {pages} {1466} (\bibinfo {year} {2012})}\BibitemShut
  {NoStop}%
\bibitem [{\citenamefont {Haberl}\ \emph {et~al.}(2016)\citenamefont {Haberl},
  \citenamefont {Strobel},\ and\ \citenamefont {Bradby}}]{Haberl2016}%
  \BibitemOpen
  \bibfield  {author} {\bibinfo {author} {\bibfnamefont {B.}~\bibnamefont
  {Haberl}}, \bibinfo {author} {\bibfnamefont {T.}~\bibnamefont {Strobel}}, \
  and\ \bibinfo {author} {\bibfnamefont {J.}~\bibnamefont {Bradby}},\
  }\href@noop {} {\bibfield  {journal} {\bibinfo  {journal} {Appl. Phys. Rev.}\
  }\textbf {\bibinfo {volume} {3}},\ \bibinfo {pages} {040808} (\bibinfo {year}
  {2016})}\BibitemShut {NoStop}%
\bibitem [{\citenamefont {Fan}\ \emph {et~al.}(2015)\citenamefont {Fan} \emph
  {et~al.}}]{Fan2015}%
  \BibitemOpen
  \bibfield  {author} {\bibinfo {author} {\bibfnamefont {Q.}~\bibnamefont
  {Fan}} \emph {et~al.},\ }\href@noop {} {\bibfield  {journal} {\bibinfo
  {journal} {Appl. Phys. Rev.}\ }\textbf {\bibinfo {volume} {118}},\ \bibinfo
  {pages} {185704} (\bibinfo {year} {2015})}\BibitemShut {NoStop}%
\bibitem [{\citenamefont {Fan}\ \emph {et~al.}(2016)\citenamefont {Fan},
  \citenamefont {Li}, \citenamefont {Peng}, \citenamefont {Meng}, \citenamefont
  {Tang},\ and\ \citenamefont {Zhong}}]{He2015}%
  \BibitemOpen
  \bibfield  {author} {\bibinfo {author} {\bibfnamefont {Q.}~\bibnamefont
  {Fan}}, \bibinfo {author} {\bibfnamefont {J.}~\bibnamefont {Li}}, \bibinfo
  {author} {\bibfnamefont {X.}~\bibnamefont {Peng}}, \bibinfo {author}
  {\bibfnamefont {L.}~\bibnamefont {Meng}}, \bibinfo {author} {\bibfnamefont
  {C.}~\bibnamefont {Tang}}, \ and\ \bibinfo {author} {\bibfnamefont
  {J.}~\bibnamefont {Zhong}},\ }\href@noop {} {\bibfield  {journal} {\bibinfo
  {journal} {Phys. Chem. Chem. Phys.}\ }\textbf {\bibinfo {volume} {18}},\
  \bibinfo {pages} {9682} (\bibinfo {year} {2016})}\BibitemShut {NoStop}%
\bibitem [{\citenamefont {Mujica}\ \emph {et~al.}(2015)\citenamefont {Mujica},
  \citenamefont {Pickard},\ and\ \citenamefont {Needs}}]{Mujica2015}%
  \BibitemOpen
  \bibfield  {author} {\bibinfo {author} {\bibfnamefont {A.}~\bibnamefont
  {Mujica}}, \bibinfo {author} {\bibfnamefont {C.~J.}\ \bibnamefont {Pickard}},
  \ and\ \bibinfo {author} {\bibfnamefont {R.}~\bibnamefont {Needs}},\
  }\href@noop {} {\bibfield  {journal} {\bibinfo  {journal} {Phys. Rev. B}\
  }\textbf {\bibinfo {volume} {91}},\ \bibinfo {pages} {214104} (\bibinfo
  {year} {2015})}\BibitemShut {NoStop}%
\bibitem [{\citenamefont {Botti}\ \emph {et~al.}(2012)\citenamefont {Botti},
  \citenamefont {Flores-Livas}, \citenamefont {Amsler}, \citenamefont
  {Goedecker},\ and\ \citenamefont {Marques}}]{Botti}%
  \BibitemOpen
  \bibfield  {author} {\bibinfo {author} {\bibfnamefont {S.}~\bibnamefont
  {Botti}}, \bibinfo {author} {\bibfnamefont {J.~A.}\ \bibnamefont
  {Flores-Livas}}, \bibinfo {author} {\bibfnamefont {M.}~\bibnamefont
  {Amsler}}, \bibinfo {author} {\bibfnamefont {S.}~\bibnamefont {Goedecker}}, \
  and\ \bibinfo {author} {\bibfnamefont {M.~A.~L.}\ \bibnamefont {Marques}},\
  }\href@noop {} {\bibfield  {journal} {\bibinfo  {journal} {Phys. Rev. B}\
  }\textbf {\bibinfo {volume} {86}},\ \bibinfo {pages} {121204} (\bibinfo
  {year} {2012})}\BibitemShut {NoStop}%
\bibitem [{\citenamefont {Kim}\ \emph {et~al.}(2015)\citenamefont {Kim},
  \citenamefont {Stefanoski}, \citenamefont {Kurakevych},\ and\ \citenamefont
  {Strobel}}]{NatMater.14.169}%
  \BibitemOpen
  \bibfield  {author} {\bibinfo {author} {\bibfnamefont {D.~Y.}\ \bibnamefont
  {Kim}}, \bibinfo {author} {\bibfnamefont {S.}~\bibnamefont {Stefanoski}},
  \bibinfo {author} {\bibfnamefont {O.~O.}\ \bibnamefont {Kurakevych}}, \ and\
  \bibinfo {author} {\bibfnamefont {T.~A.}\ \bibnamefont {Strobel}},\
  }\href@noop {} {\bibfield  {journal} {\bibinfo  {journal} {Nat. Mater.}\
  }\textbf {\bibinfo {volume} {14}},\ \bibinfo {pages} {169} (\bibinfo {year}
  {2015})}\BibitemShut {NoStop}%
\bibitem [{\citenamefont {Raffy}\ \emph {et~al.}(2002)\citenamefont {Raffy},
  \citenamefont {Furthm{\"u}ller},\ and\ \citenamefont {Bechstedt}}]{34}%
  \BibitemOpen
  \bibfield  {author} {\bibinfo {author} {\bibfnamefont {C.}~\bibnamefont
  {Raffy}}, \bibinfo {author} {\bibfnamefont {J.}~\bibnamefont
  {Furthm{\"u}ller}}, \ and\ \bibinfo {author} {\bibfnamefont {F.}~\bibnamefont
  {Bechstedt}},\ }\href@noop {} {\bibfield  {journal} {\bibinfo  {journal}
  {Phys. Rev. B}\ }\textbf {\bibinfo {volume} {66}},\ \bibinfo {pages} {075201}
  (\bibinfo {year} {2002})}\BibitemShut {NoStop}%
\bibitem [{\citenamefont {Wen}\ \emph {et~al.}(2008)\citenamefont {Wen},
  \citenamefont {Zhao}, \citenamefont {Bucknum}, \citenamefont {Yao},\ and\
  \citenamefont {Li}}]{35}%
  \BibitemOpen
  \bibfield  {author} {\bibinfo {author} {\bibfnamefont {B.}~\bibnamefont
  {Wen}}, \bibinfo {author} {\bibfnamefont {J.}~\bibnamefont {Zhao}}, \bibinfo
  {author} {\bibfnamefont {M.~J.}\ \bibnamefont {Bucknum}}, \bibinfo {author}
  {\bibfnamefont {P.}~\bibnamefont {Yao}}, \ and\ \bibinfo {author}
  {\bibfnamefont {T.}~\bibnamefont {Li}},\ }\href@noop {} {\bibfield  {journal}
  {\bibinfo  {journal} {Diamond Relat. Mater.}\ }\textbf {\bibinfo {volume}
  {17}},\ \bibinfo {pages} {356} (\bibinfo {year} {2008})}\BibitemShut
  {NoStop}%
\bibitem [{SAC()}]{SACADA1}%
  \BibitemOpen
  \href@noop {} {}\bibinfo {note} {\texttt{http://sacada.sctms.ru}}\BibitemShut
  {NoStop}%
\bibitem [{\citenamefont {Hoffmann}\ \emph {et~al.}(2016)\citenamefont
  {Hoffmann}, \citenamefont {Kabanov}, \citenamefont {Golov},\ and\
  \citenamefont {Proserpio}}]{SACADA2}%
  \BibitemOpen
  \bibfield  {author} {\bibinfo {author} {\bibfnamefont {R.}~\bibnamefont
  {Hoffmann}}, \bibinfo {author} {\bibfnamefont {A.~A.}\ \bibnamefont
  {Kabanov}}, \bibinfo {author} {\bibfnamefont {A.~A.}\ \bibnamefont {Golov}},
  \ and\ \bibinfo {author} {\bibfnamefont {D.~M.}\ \bibnamefont {Proserpio}},\
  }\href@noop {} {\bibfield  {journal} {\bibinfo  {journal} {Angew. Chem. Int.
  Ed.}\ }\textbf {\bibinfo {volume} {55}},\ \bibinfo {pages} {10962} (\bibinfo
  {year} {2016})}\BibitemShut {NoStop}%
\bibitem [{\citenamefont {Dovesi}\ \emph
  {et~al.}(2014{\natexlab{a}})\citenamefont {Dovesi} \emph
  {et~al.}}]{CRYSTAL1}%
  \BibitemOpen
  \bibfield  {author} {\bibinfo {author} {\bibfnamefont {R.}~\bibnamefont
  {Dovesi}} \emph {et~al.},\ }\href@noop {} {\bibfield  {journal} {\bibinfo
  {journal} {Int. J. Quantum Chem.}\ }\textbf {\bibinfo {volume} {114}},\
  \bibinfo {pages} {1287} (\bibinfo {year} {2014}{\natexlab{a}})}\BibitemShut
  {NoStop}%
\bibitem [{\citenamefont {Dovesi}\ \emph
  {et~al.}(2014{\natexlab{b}})\citenamefont {Dovesi} \emph
  {et~al.}}]{CRYSTAL2}%
  \BibitemOpen
  \bibfield  {author} {\bibinfo {author} {\bibfnamefont {R.}~\bibnamefont
  {Dovesi}} \emph {et~al.},\ }\href@noop {} {\emph {\bibinfo {title} {CRYSTAL14
  User's Manual}}},\ \bibinfo {organization} {Univ. of Torino},\ \bibinfo
  {address} {Torino, Italy} (\bibinfo {year} {2014}{\natexlab{b}})\BibitemShut
  {NoStop}%
\bibitem [{\citenamefont {Perdew}\ \emph {et~al.}(1996)\citenamefont {Perdew},
  \citenamefont {Burke},\ and\ \citenamefont {Ernzerhof}}]{PBE}%
  \BibitemOpen
  \bibfield  {author} {\bibinfo {author} {\bibfnamefont {J.~P.}\ \bibnamefont
  {Perdew}}, \bibinfo {author} {\bibfnamefont {K.}~\bibnamefont {Burke}}, \
  and\ \bibinfo {author} {\bibfnamefont {M.}~\bibnamefont {Ernzerhof}},\
  }\href@noop {} {\bibfield  {journal} {\bibinfo  {journal} {Phys. Rev. Lett.}\
  }\textbf {\bibinfo {volume} {77}},\ \bibinfo {pages} {3865} (\bibinfo {year}
  {1996})}\BibitemShut {NoStop}%
\bibitem [{\citenamefont {Peitinger}\ \emph {et~al.}(2013)\citenamefont
  {Peitinger}, \citenamefont {Oliveira},\ and\ \citenamefont {Bredow}}]{Basis}%
  \BibitemOpen
  \bibfield  {author} {\bibinfo {author} {\bibfnamefont {M.~F.}\ \bibnamefont
  {Peitinger}}, \bibinfo {author} {\bibfnamefont {D.~V.}\ \bibnamefont
  {Oliveira}}, \ and\ \bibinfo {author} {\bibfnamefont {T.}~\bibnamefont
  {Bredow}},\ }\href@noop {} {\bibfield  {journal} {\bibinfo  {journal} {J.
  Comput. Chem.}\ }\textbf {\bibinfo {volume} {34}},\ \bibinfo {pages} {451}
  (\bibinfo {year} {2013})}\BibitemShut {NoStop}%
\bibitem [{\citenamefont {Monkhorst}\ and\ \citenamefont
  {Pack}(1976)}]{MPpoints}%
  \BibitemOpen
  \bibfield  {author} {\bibinfo {author} {\bibfnamefont {H.~J.}\ \bibnamefont
  {Monkhorst}}\ and\ \bibinfo {author} {\bibfnamefont {J.~D.}\ \bibnamefont
  {Pack}},\ }\href@noop {} {\bibfield  {journal} {\bibinfo  {journal} {Phys.
  Rev. B}\ }\textbf {\bibinfo {volume} {13}},\ \bibinfo {pages} {5188}
  (\bibinfo {year} {1976})}\BibitemShut {NoStop}%
\bibitem [{\citenamefont {Kresse}\ and\ \citenamefont
  {Furthm{\"u}ller}(1996)}]{VASP1}%
  \BibitemOpen
  \bibfield  {author} {\bibinfo {author} {\bibfnamefont {G.}~\bibnamefont
  {Kresse}}\ and\ \bibinfo {author} {\bibfnamefont {J.}~\bibnamefont
  {Furthm{\"u}ller}},\ }\href@noop {} {\bibfield  {journal} {\bibinfo
  {journal} {Phys. Rev. B}\ }\textbf {\bibinfo {volume} {54}},\ \bibinfo
  {pages} {11169} (\bibinfo {year} {1996})}\BibitemShut {NoStop}%
\bibitem [{\citenamefont {Kresse}\ and\ \citenamefont {Joubert}(1999)}]{VASP2}%
  \BibitemOpen
  \bibfield  {author} {\bibinfo {author} {\bibfnamefont {G.}~\bibnamefont
  {Kresse}}\ and\ \bibinfo {author} {\bibfnamefont {D.}~\bibnamefont
  {Joubert}},\ }\href@noop {} {\bibfield  {journal} {\bibinfo  {journal} {Phys.
  Rev. B}\ }\textbf {\bibinfo {volume} {59}},\ \bibinfo {pages} {1758}
  (\bibinfo {year} {1999})}\BibitemShut {NoStop}%
\bibitem [{\citenamefont {Heyd}\ \emph {et~al.}(2006)\citenamefont {Heyd},
  \citenamefont {Scuseria},\ and\ \citenamefont {Ernzerhof}}]{HSE06}%
  \BibitemOpen
  \bibfield  {author} {\bibinfo {author} {\bibfnamefont {J.}~\bibnamefont
  {Heyd}}, \bibinfo {author} {\bibfnamefont {G.}~\bibnamefont {Scuseria}}, \
  and\ \bibinfo {author} {\bibfnamefont {M.}~\bibnamefont {Ernzerhof}},\
  }\href@noop {} {\bibfield  {journal} {\bibinfo  {journal} {J. Chem. Phys.}\
  }\textbf {\bibinfo {volume} {124}},\ \bibinfo {pages} {219906} (\bibinfo
  {year} {2006})}\BibitemShut {NoStop}%
\bibitem [{\citenamefont {Birch}(1947)}]{BM}%
  \BibitemOpen
  \bibfield  {author} {\bibinfo {author} {\bibfnamefont {F.}~\bibnamefont
  {Birch}},\ }\href@noop {} {\bibfield  {journal} {\bibinfo  {journal} {Phys.
  Rev.}\ }\textbf {\bibinfo {volume} {71}},\ \bibinfo {pages} {809} (\bibinfo
  {year} {1947})}\BibitemShut {NoStop}%
\bibitem [{\citenamefont {Pascale}\ \emph {et~al.}(2004)\citenamefont
  {Pascale}, \citenamefont {Zicovich-Wilson}, \citenamefont {Lopez},
  \citenamefont {Civalleri}, \citenamefont {Orlando},\ and\ \citenamefont
  {Dovesi}}]{thermodynamics}%
  \BibitemOpen
  \bibfield  {author} {\bibinfo {author} {\bibfnamefont {F.}~\bibnamefont
  {Pascale}}, \bibinfo {author} {\bibfnamefont {C.}~\bibnamefont
  {Zicovich-Wilson}}, \bibinfo {author} {\bibfnamefont {F.}~\bibnamefont
  {Lopez}}, \bibinfo {author} {\bibfnamefont {B.}~\bibnamefont {Civalleri}},
  \bibinfo {author} {\bibfnamefont {R.}~\bibnamefont {Orlando}}, \ and\
  \bibinfo {author} {\bibfnamefont {R.}~\bibnamefont {Dovesi}},\ }\href@noop {}
  {\bibfield  {journal} {\bibinfo  {journal} {J. Comput. Chem.}\ }\textbf
  {\bibinfo {volume} {25}},\ \bibinfo {pages} {888} (\bibinfo {year}
  {2004})}\BibitemShut {NoStop}%
\bibitem [{\citenamefont {Haynes}(2015)}]{CRC96th}%
  \BibitemOpen
  \bibfield  {author} {\bibinfo {author} {\bibfnamefont {W.~M.}\ \bibnamefont
  {Haynes}},\ }\href@noop {} {\emph {\bibinfo {title} {CRC Handbook of
  Chemistry and Physics 2015-2016}}},\ \bibinfo {edition} {96th}\ ed.\
  (\bibinfo  {publisher} {CRC Press/Taylor and Francis},\ \bibinfo {address}
  {Boca Raton, FL, USA},\ \bibinfo {year} {2015})\BibitemShut {NoStop}%
\bibitem [{\citenamefont {Born}\ and\ \citenamefont
  {Huang}(1954)}]{BornHuang1954}%
  \BibitemOpen
  \bibfield  {author} {\bibinfo {author} {\bibfnamefont {M.}~\bibnamefont
  {Born}}\ and\ \bibinfo {author} {\bibfnamefont {K.}~\bibnamefont {Huang}},\
  }\href@noop {} {\emph {\bibinfo {title} {Dynamical Theory of Crystal
  Lattices}}}\ (\bibinfo  {publisher} {Oxford University Press},\ \bibinfo
  {address} {Oxford},\ \bibinfo {year} {1954})\BibitemShut {NoStop}%
\bibitem [{\citenamefont {Mouhat}\ and\ \citenamefont {Coudert}(2014)}]{Cab}%
  \BibitemOpen
  \bibfield  {author} {\bibinfo {author} {\bibfnamefont {F.}~\bibnamefont
  {Mouhat}}\ and\ \bibinfo {author} {\bibfnamefont {F.-X.}\ \bibnamefont
  {Coudert}},\ }\href@noop {} {\bibfield  {journal} {\bibinfo  {journal} {Phys.
  Rev. B}\ }\textbf {\bibinfo {volume} {90}},\ \bibinfo {pages} {224104}
  (\bibinfo {year} {2014})}\BibitemShut {NoStop}%
\bibitem [{\citenamefont {Gilman}(2003)}]{Gilman2003}%
  \BibitemOpen
  \bibfield  {author} {\bibinfo {author} {\bibfnamefont {J.~J.}\ \bibnamefont
  {Gilman}},\ }\href@noop {} {\emph {\bibinfo {title} {Electronic Basis of the
  Strength of Materials}}}\ (\bibinfo  {publisher} {Cambridge University
  Press},\ \bibinfo {address} {Cambridge},\ \bibinfo {year} {2003})\BibitemShut
  {NoStop}%
\bibitem [{\citenamefont {Pugh}(1954)}]{Pugh}%
  \BibitemOpen
  \bibfield  {author} {\bibinfo {author} {\bibfnamefont {S.}~\bibnamefont
  {Pugh}},\ }\href@noop {} {\bibfield  {journal} {\bibinfo  {journal} {Philos.
  Mag.}\ }\textbf {\bibinfo {volume} {45}},\ \bibinfo {pages} {823} (\bibinfo
  {year} {1954})}\BibitemShut {NoStop}%
\bibitem [{\citenamefont {Long}\ \emph {et~al.}(2016)\citenamefont {Long},
  \citenamefont {Nie}, \citenamefont {Shang}, \citenamefont {Wang},
  \citenamefont {Du}, \citenamefont {Jin},\ and\ \citenamefont {Liu}}]{QLong}%
  \BibitemOpen
  \bibfield  {author} {\bibinfo {author} {\bibfnamefont {Q.}~\bibnamefont
  {Long}}, \bibinfo {author} {\bibfnamefont {X.}~\bibnamefont {Nie}}, \bibinfo
  {author} {\bibfnamefont {S.-L.}\ \bibnamefont {Shang}}, \bibinfo {author}
  {\bibfnamefont {J.}~\bibnamefont {Wang}}, \bibinfo {author} {\bibfnamefont
  {Y.}~\bibnamefont {Du}}, \bibinfo {author} {\bibfnamefont {Z.}~\bibnamefont
  {Jin}}, \ and\ \bibinfo {author} {\bibfnamefont {Z.-K.}\ \bibnamefont
  {Liu}},\ }\href@noop {} {\bibfield  {journal} {\bibinfo  {journal} {Comput.
  Mater. Sci.}\ }\textbf {\bibinfo {volume} {121}},\ \bibinfo {pages} {167}
  (\bibinfo {year} {2016})}\BibitemShut {NoStop}%
\bibitem [{\citenamefont {Voigt}(1928)}]{Voigt}%
  \BibitemOpen
  \bibfield  {author} {\bibinfo {author} {\bibfnamefont {W.}~\bibnamefont
  {Voigt}},\ }\href@noop {} {\emph {\bibinfo {title} {Lehrburch der
  Kristallphysik}}}\ (\bibinfo  {publisher} {Teubner},\ \bibinfo {address}
  {Leipzig},\ \bibinfo {year} {1928})\BibitemShut {NoStop}%
\bibitem [{\citenamefont {Becke}\ and\ \citenamefont {Edgecombe}(1990)}]{ELF1}%
  \BibitemOpen
  \bibfield  {author} {\bibinfo {author} {\bibfnamefont {A.~D.}\ \bibnamefont
  {Becke}}\ and\ \bibinfo {author} {\bibfnamefont {K.~E.}\ \bibnamefont
  {Edgecombe}},\ }\href@noop {} {\bibfield  {journal} {\bibinfo  {journal} {J.
  Chem. Phys.}\ }\textbf {\bibinfo {volume} {92}},\ \bibinfo {pages}
  {5397–5403} (\bibinfo {year} {1990})}\BibitemShut {NoStop}%
\bibitem [{ELF(1992)}]{ELF2}%
  \BibitemOpen
  \href@noop {} {\bibfield  {journal} {\bibinfo  {journal} {Angew. Chem. Int.
  Ed.}\ }\textbf {\bibinfo {volume} {31}},\ \bibinfo {pages} {187–188}
  (\bibinfo {year} {1992})}\BibitemShut {NoStop}%
\bibitem [{\citenamefont {Kleinman}(1962)}]{Kleiman}%
  \BibitemOpen
  \bibfield  {author} {\bibinfo {author} {\bibfnamefont {L.}~\bibnamefont
  {Kleinman}},\ }\href@noop {} {\bibfield  {journal} {\bibinfo  {journal}
  {Phys. Rev.}\ }\textbf {\bibinfo {volume} {128}},\ \bibinfo {pages} {2614}
  (\bibinfo {year} {1962})}\BibitemShut {NoStop}%
\bibitem [{\citenamefont {Baima}\ \emph {et~al.}(2016)\citenamefont {Baima},
  \citenamefont {Zelferino}, \citenamefont {Olivero}, \citenamefont {Erba},\
  and\ \citenamefont {Dovesi}}]{Baima2016}%
  \BibitemOpen
  \bibfield  {author} {\bibinfo {author} {\bibfnamefont {J.}~\bibnamefont
  {Baima}}, \bibinfo {author} {\bibfnamefont {A.}~\bibnamefont {Zelferino}},
  \bibinfo {author} {\bibfnamefont {P.}~\bibnamefont {Olivero}}, \bibinfo
  {author} {\bibfnamefont {A.}~\bibnamefont {Erba}}, \ and\ \bibinfo {author}
  {\bibfnamefont {R.}~\bibnamefont {Dovesi}},\ }\href {\doibase
  10.1039/C5CP06672G} {\bibfield  {journal} {\bibinfo  {journal} {Phys. Chem.
  Chem. Phys.}\ } (\bibinfo {year} {2016}),\ 10.1039/C5CP06672G}\BibitemShut
  {NoStop}%
\bibitem [{\citenamefont {De}\ and\ \citenamefont {Pryor}(2014)}]{De}%
  \BibitemOpen
  \bibfield  {author} {\bibinfo {author} {\bibfnamefont {A.}~\bibnamefont
  {De}}\ and\ \bibinfo {author} {\bibfnamefont {C.~E.}\ \bibnamefont {Pryor}},\
  }\href@noop {} {\bibfield  {journal} {\bibinfo  {journal} {J. Phys. Condens.
  Matter}\ }\textbf {\bibinfo {volume} {26}},\ \bibinfo {pages} {045801}
  (\bibinfo {year} {2014})}\BibitemShut {NoStop}%
\bibitem [{\citenamefont {Hafner}\ \emph {et~al.}(1988)\citenamefont {Hafner},
  \citenamefont {Tegze},\ and\ \citenamefont {Jank}}]{JLessCommonMet.145.531}%
  \BibitemOpen
  \bibfield  {author} {\bibinfo {author} {\bibfnamefont {J.}~\bibnamefont
  {Hafner}}, \bibinfo {author} {\bibfnamefont {M.}~\bibnamefont {Tegze}}, \
  and\ \bibinfo {author} {\bibfnamefont {W.}~\bibnamefont {Jank}},\ }\href@noop
  {} {\bibfield  {journal} {\bibinfo  {journal} {J. Less Common Met.}\ }\textbf
  {\bibinfo {volume} {145}},\ \bibinfo {pages} {531} (\bibinfo {year}
  {1988})}\BibitemShut {NoStop}%
\bibitem [{\citenamefont {Shimomura}\ \emph {et~al.}(1974)\citenamefont
  {Shimomura}, \citenamefont {Minomura}, \citenamefont {Sakai}, \citenamefont
  {Asaumi}, \citenamefont {Tamura}, \citenamefont {Fukushima},\ and\
  \citenamefont {Endo}}]{PhilosMag.29.547}%
  \BibitemOpen
  \bibfield  {author} {\bibinfo {author} {\bibfnamefont {O.}~\bibnamefont
  {Shimomura}}, \bibinfo {author} {\bibfnamefont {S.}~\bibnamefont {Minomura}},
  \bibinfo {author} {\bibfnamefont {N.}~\bibnamefont {Sakai}}, \bibinfo
  {author} {\bibfnamefont {K.}~\bibnamefont {Asaumi}}, \bibinfo {author}
  {\bibfnamefont {K.}~\bibnamefont {Tamura}}, \bibinfo {author} {\bibfnamefont
  {J.}~\bibnamefont {Fukushima}}, \ and\ \bibinfo {author} {\bibfnamefont
  {H.}~\bibnamefont {Endo}},\ }\href@noop {} {\bibfield  {journal} {\bibinfo
  {journal} {Philos. Mag.}\ }\textbf {\bibinfo {volume} {29}},\ \bibinfo
  {pages} {547} (\bibinfo {year} {1974})}\BibitemShut {NoStop}%
\bibitem [{\citenamefont {Hummer}\ \emph {et~al.}(2009)\citenamefont {Hummer},
  \citenamefont {Harl},\ and\ \citenamefont {G.}}]{Hummer2009}%
  \BibitemOpen
  \bibfield  {author} {\bibinfo {author} {\bibfnamefont {K.}~\bibnamefont
  {Hummer}}, \bibinfo {author} {\bibfnamefont {J.}~\bibnamefont {Harl}}, \ and\
  \bibinfo {author} {\bibfnamefont {K.}~\bibnamefont {G.}},\ }\href@noop {}
  {\bibfield  {journal} {\bibinfo  {journal} {Phys. Rev. B}\ }\textbf {\bibinfo
  {volume} {80}},\ \bibinfo {pages} {115205} (\bibinfo {year}
  {2009})}\BibitemShut {NoStop}%
\bibitem [{Sad(1999)}]{SadaoAdachi}%
  \BibitemOpen
  in\ \href@noop {} {\emph {\bibinfo {booktitle} {Optical Constants of
  Crystalline and Amorphous Semiconductors}}},\ \bibinfo {editor} {edited by\
  \bibinfo {editor} {\bibfnamefont {A.}~\bibnamefont {Sadao}}}\ (\bibinfo
  {publisher} {Springer},\ \bibinfo {address} {USA},\ \bibinfo {year} {1999})\
  \bibinfo {edition} {714th}\ ed.\BibitemShut {Stop}%
\bibitem [{\citenamefont {Ferlauto}\ \emph {et~al.}(2002)\citenamefont
  {Ferlauto}, \citenamefont {Ferreira}, \citenamefont {Pearce}, \citenamefont
  {Wronski}, \citenamefont {Collins}, \citenamefont {Deng},\ and\ \citenamefont
  {Ganguly}}]{FerlautoAmorph}%
  \BibitemOpen
  \bibfield  {author} {\bibinfo {author} {\bibfnamefont {A.~S.}\ \bibnamefont
  {Ferlauto}}, \bibinfo {author} {\bibfnamefont {G.~M.}\ \bibnamefont
  {Ferreira}}, \bibinfo {author} {\bibfnamefont {J.~M.}\ \bibnamefont
  {Pearce}}, \bibinfo {author} {\bibfnamefont {C.~R.}\ \bibnamefont {Wronski}},
  \bibinfo {author} {\bibfnamefont {R.}~\bibnamefont {Collins}}, \bibinfo
  {author} {\bibfnamefont {X.}~\bibnamefont {Deng}}, \ and\ \bibinfo {author}
  {\bibfnamefont {G.}~\bibnamefont {Ganguly}},\ }\href@noop {} {\bibfield
  {journal} {\bibinfo  {journal} {J. Appl. Phys.}\ }\textbf {\bibinfo {volume}
  {92}},\ \bibinfo {pages} {2424} (\bibinfo {year} {2002})}\BibitemShut
  {NoStop}%
\end{thebibliography}%

\newpage
\begin{table}
\caption{\label{Table:1} Comparison, for the diamond structure
($Fd\bar 3m$), between the experimental and computed (CRYSTAL) values of
some basic properties: cell parameters, density, bulk modulus,
pressure derivative of the bulk modulus, dielectric constant, indirect and
direct (at the $\Gamma$ point) band gaps.}

\begin{tabular}{|c|c|c|c|c|c|c|c|}
\hline
 $Fd\bar 3m$    & $a, {\AA}$   & $\rho$, \si{kg.m^{-3}}&    $B$, GPa & $B'$ &
$\varepsilon$ &
$\triangle E_{gap}$, eV & $\triangle E_{\Gamma}$, eV \\
\hline

Si, exp.& 5.431&   2329&   98& &      11.7&    1.12 &   3.4
\\ \hline
Si, PBE/HSE06& 5.414&   2342&   92&  4.63&    11.9&    1.21/1.67&    2.9/3.5 \\
\hline


Ge, exp.& 5.658&   5323&   75& &      16.0&    0.66&    0.8\\
\hline

Ge, PBE/HSE06 & 5.685& 5198&   67&  4.86&    16.2&    0.52/1.05&    0.7/1.24 \\
\hline


\end{tabular}
\end{table}

\begin{table}
\caption{\label{Table:2} The density, difference in energy from the
corresponding diamond phase at zero pressure, indirect and direct (at $\Gamma$)
band gaps, bulk modulus, pressure derivative of the bulk
modulus, shear modulus, and static dielectric constants for the various
structures of silicon.}

\begin{tabular}{|c|c|c|c|c|c|c|}
\hline $\#N$, Sym.~Group & $\#88$,$Pnma$ &$\#50$,$Pnma$& $\#55$,$Pmma$&
$\#26$,$P2/m$ &
$\#27$,$C2/m$ & $\#28$,$Pbam$\\
\hline $\rho$, \si{kg.m^{-3}} & 2293&   2291&   2305 &  2298 &
2291& 2271\\
\hline $\triangle E$, eV & 0.037&   0.036&   0.035&   0.039& 0.047&
0.036\\ \hline $E_{gap}$, eV, PBE/HSE06 &    1.46/2.03& 1.42/1.97&
1.26/1.78& 0.97/1.51&
1.02/1.29& 1.31/1.87\\
\hline $\triangle E_\Gamma$, eV,PBE/HSE06  &1.56/2.12& 1.51/2.08&
1.58/2.10& 1.10/1.66&
1.12/1.60& 1.57/2.12\\
\hline $B$, GPa & 78&  84&  83&  84&  84&  86\\ \hline $B'$&  2.74
&4.31& 3.75 &   4.14&    4.35&    4.64\\
\hline $G$, GPa & 48&  48&  48&  50&  51& 51 \\
\hline $\varepsilon_{xx}$, $\varepsilon_{yy}$, $\varepsilon_{zz}$ &
{11.9,12.1,12.1} & 11.9,12.1,12.5& 11.7,11.4,11.9& 11.9,11.8,12.2&
12.7,11.9,12.1& 12.6,11.5,11.9\\
\hline
\end{tabular}
\end{table}

\begin{table}
\caption{\label{Table:3} Same properties as in Table~\ref{Table:2} for the
equivalent germanium allotropes}

\begin{tabular}{|c|c|c|c|c|c|c|}
\hline $\#N$, Sym.~Group & $\#88$,$Pnma$ &$\#50$,$Pnma$& $\#55$,$Pmma$&
$\#26$,$P2/m$ &
$\#27$,$C2/m$ & $\#28$,$Pbam$\\
\hline $\rho$, \si{kg.m^{-3}}    &5251&   5086&   5121&   5102&
5087& 5054 \\
\hline $\Delta E$, eV&  0.037&   0.038&   0.036&   0.044&   0.052&   0.033\\
\hline $E_{gap}$, eV, PBE/HSE06   & 0.87/1.46&    0.86/1.43&
0.61/1.10& 0.42/0.86& 0.23/0.93&
0.65/1.23\\
\hline $\Delta E_\Gamma$, eV, PBE/HSE06 & 0.95/1.54& 0.96/1.45&
1.08/1.59& 0.42/0.86&
0.60/1.26 &0.65/1.33\\
\hline $B$, GPa & 57 & 59&  59&  59&  59&  60\\
\hline $B'$ &4.24 &   4.54&    4.38&    4.44&    4.53&    4.61\\
\hline $G$, GPa & 43&  43&  44&  45&  45&  46\\
\hline $\varepsilon_{xx}$, $\varepsilon_{yy}$, $\varepsilon_{zz}$
&14.9,16.5,15.7&  15.7,16.4,16.2& 14.8,15.3,15.5& 15.6,15.9,15.5&
16.7,17.8,16.8& 16.1,14.9,15.5\\
\hline
\end{tabular}
\end{table}

\begin{table}
\caption{\label{Table:4} Bond lengths and angles, and shortest
non-bonding distances in Si and Ge allotropes.}

\begin{tabular}{|c|c|c|c|}
\hline Structure &  Bond lengths, $\AA$&  Bond angles, degrees &
\begin{tabular}{cc} Shortest non-bonding\\ distance, ${\AA}$\end{tabular}\\
\hline Si$\#26$& 2.308-2.382& 95.98-126.34&    3.818\\
\hline Si$\#27$& 2.308-2.415& 95.31-125.84&    3.820\\
\hline Si$\#28$& 2.308-2.406& 93.14-124.12&    3.818\\
\hline Si$\#50$& 2.301-2.386& 96.95-127.46&    3.826\\
\hline Si$\#55$& 2.322-2.422& 97.91-120.04&    3.828\\
\hline Si$\#88$& 2.322-2.380& 98.16-119.32&    3.837\\
\hline Ge$\#26$& 2.436-2.562& 94.76-126.16&    4.055\\
\hline Ge$\#27$& 2.436-2.533& 94.76-126.16&    4.055\\
\hline Ge$\#28$& 2.447-2.539& 92.88-120.79&    4.057\\
\hline Ge$\#50$& 2.439-2.530& 96.29-122.11&    4.067\\
\hline Ge$\#55$& 2.451-2.547& 98.65-119.12&    4.058\\
\hline Ge$\#88$& 2.459-2.514& 97.51-120.61&    4.068\\
\hline
\end{tabular}
\end{table}

\newpage

\begin{figure}[H]
\includegraphics[width=1.0\textwidth, clip=]{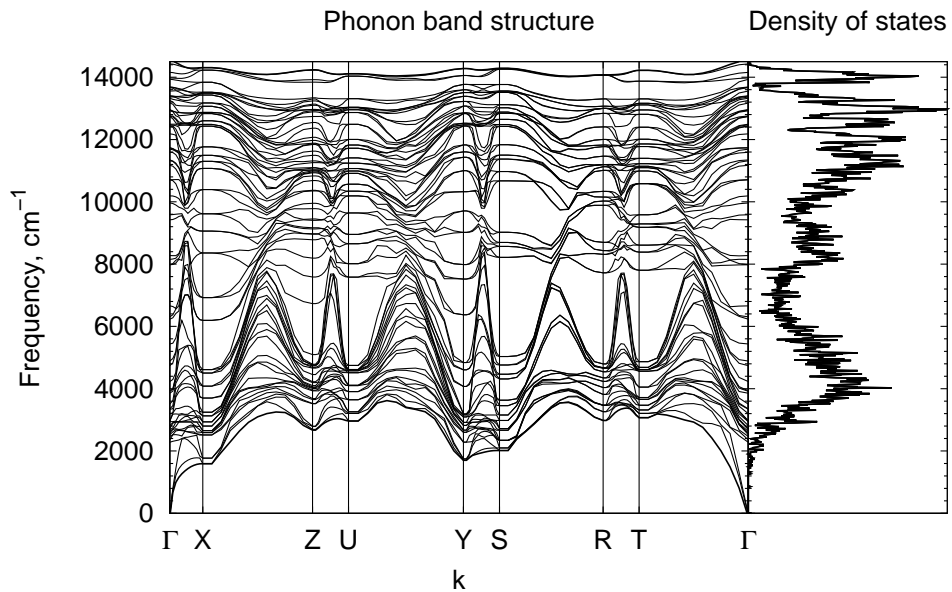}
\caption{Phonon band structure and DOS for silicon allotrope Si$\#28$.
}\label{fig:PDOSSi}
\end{figure}

\begin{figure}[H]
\includegraphics[width=1.0\textwidth, clip=]{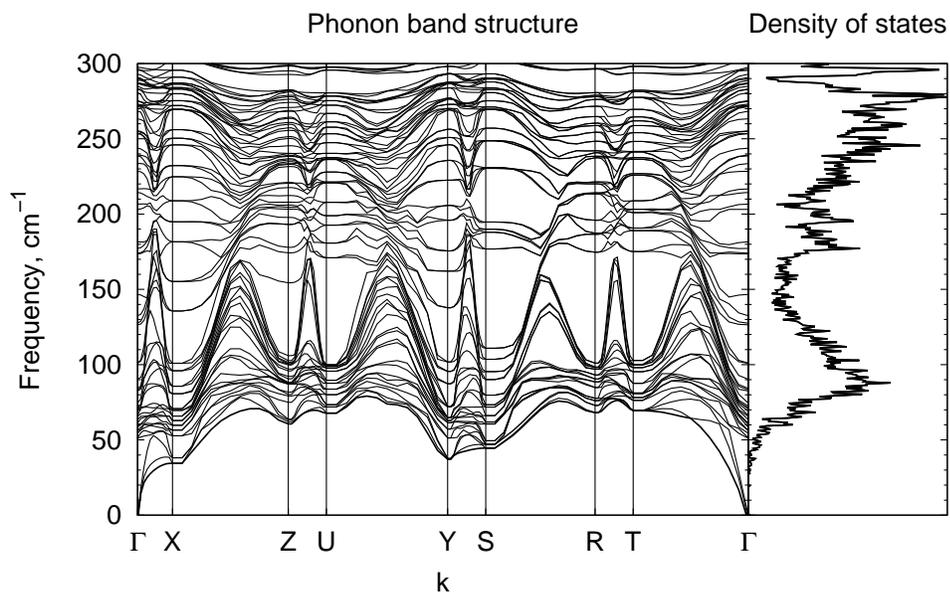}
\caption{Phonon band structure and DOS for germanium allotrope
Ge$\#28$.}\label{fig:PDOSGe}
\end{figure}

\begin{figure}[H]
\includegraphics[width=0.5\textwidth, clip=]{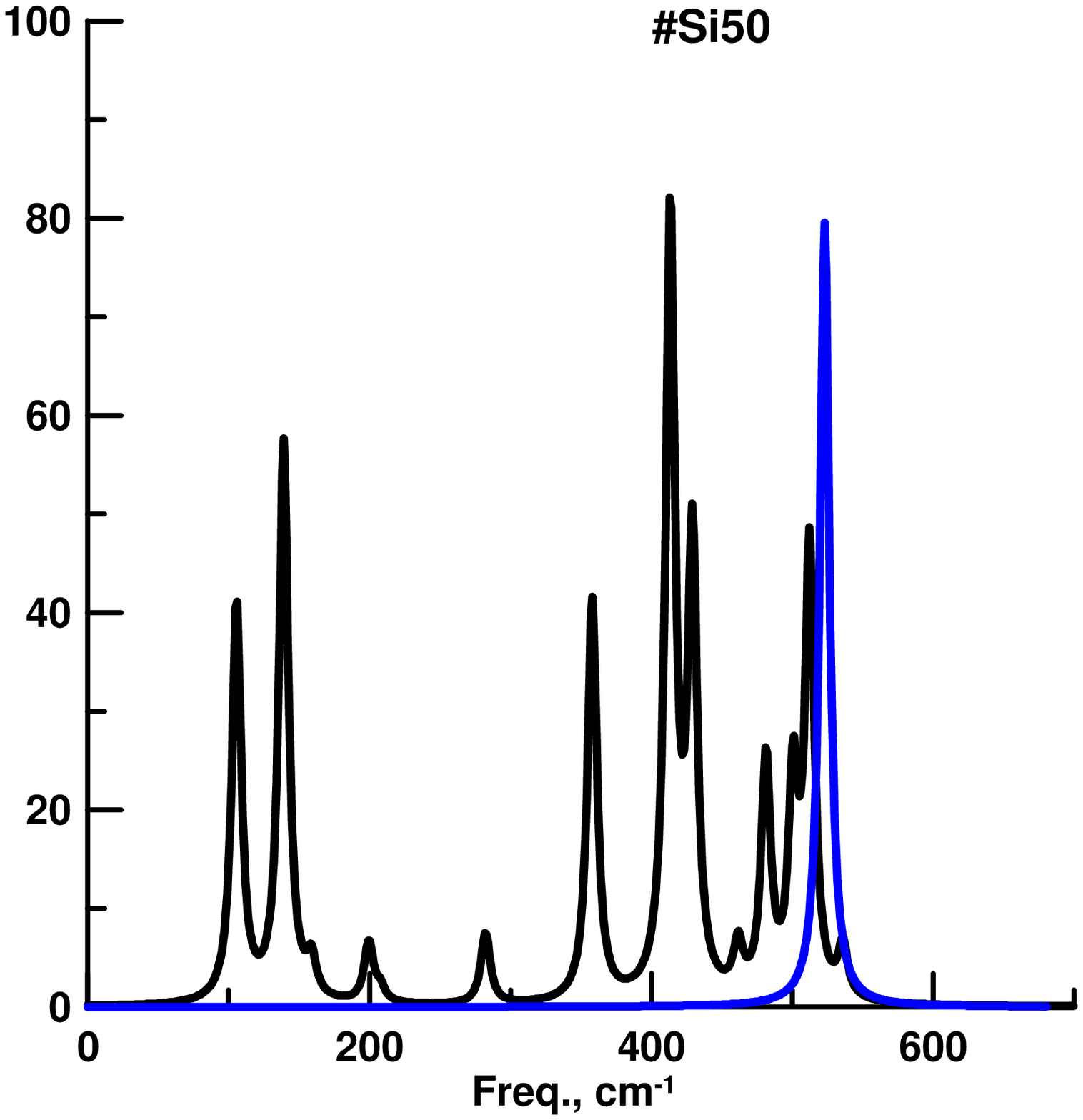}\includegraphics[width=0.5\textwidth, clip=]{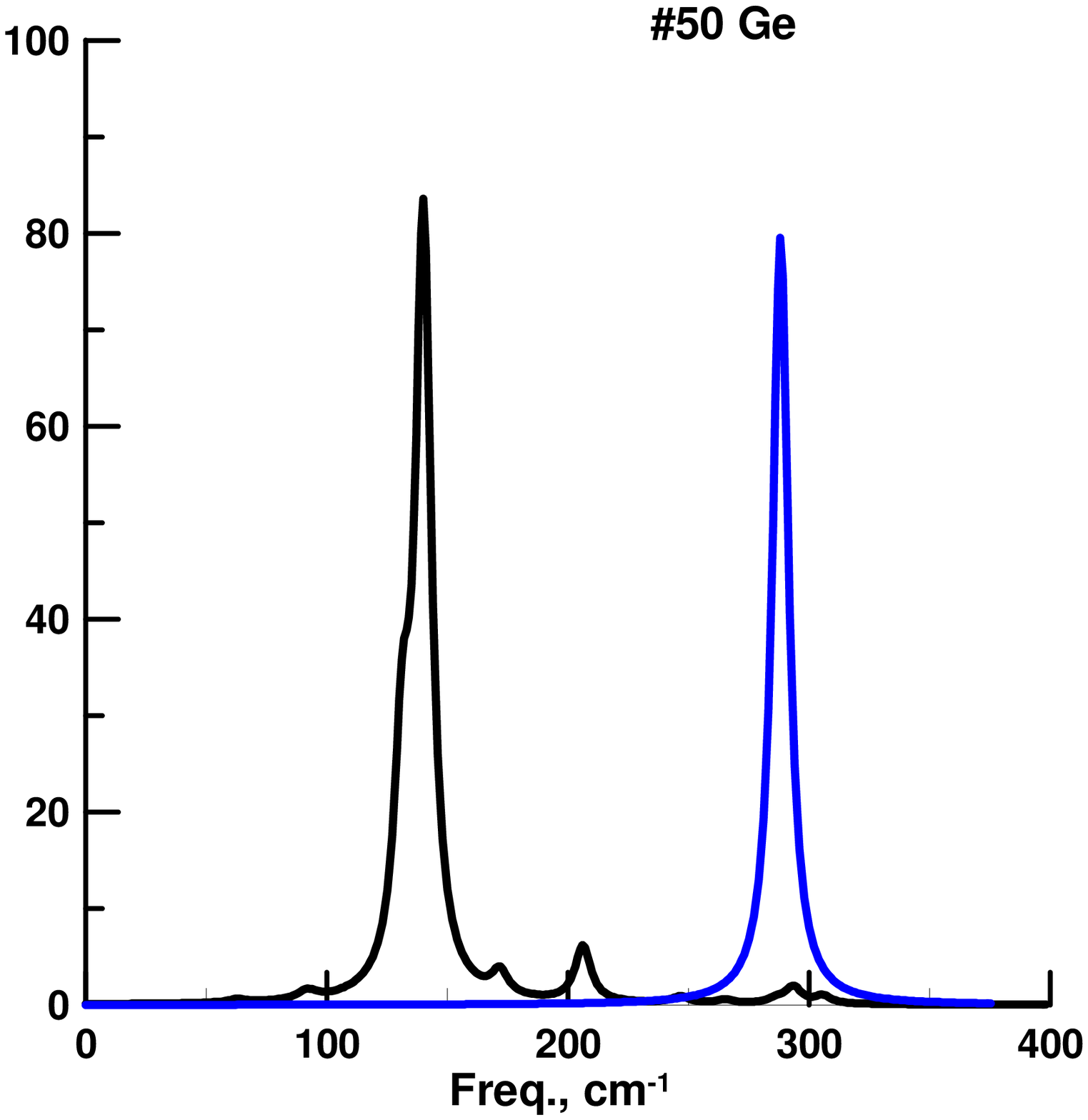}
\caption{Raman shift spectra for silicon (left panel) and germanium
(right panel) allotropes $\#50$ (black lines) and diamond structures
(blue lines).}\label{fig:RamSiGe}
\end{figure}

\begin{figure}[H]
\includegraphics[width=1.0\textwidth, clip=]{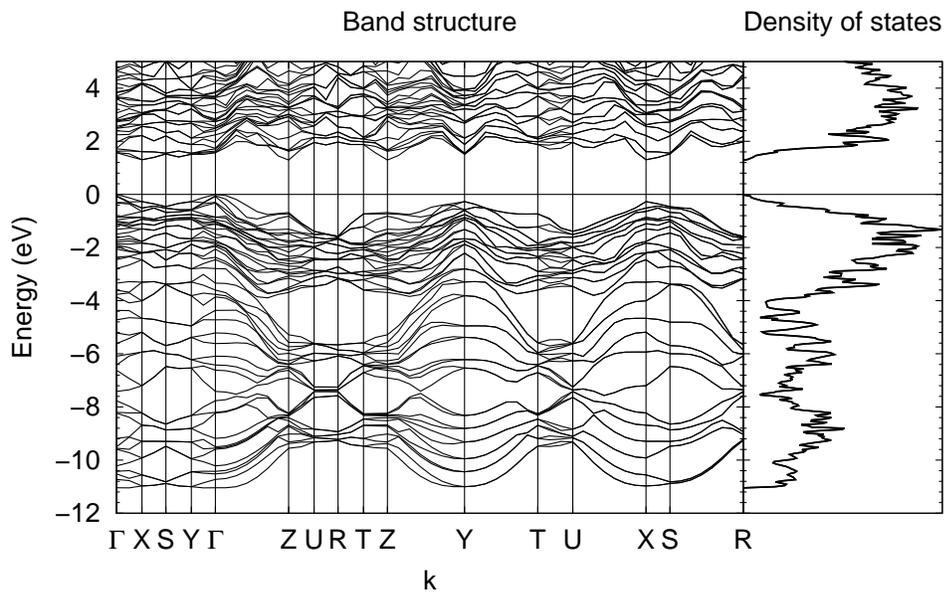}
\caption{Electronic band structure and DOS for silicon allotrope
Si$\#28$.}\label{fig:BandSi}
\end{figure}

\begin{figure}[H]
\includegraphics[width=1.0\textwidth, clip=]{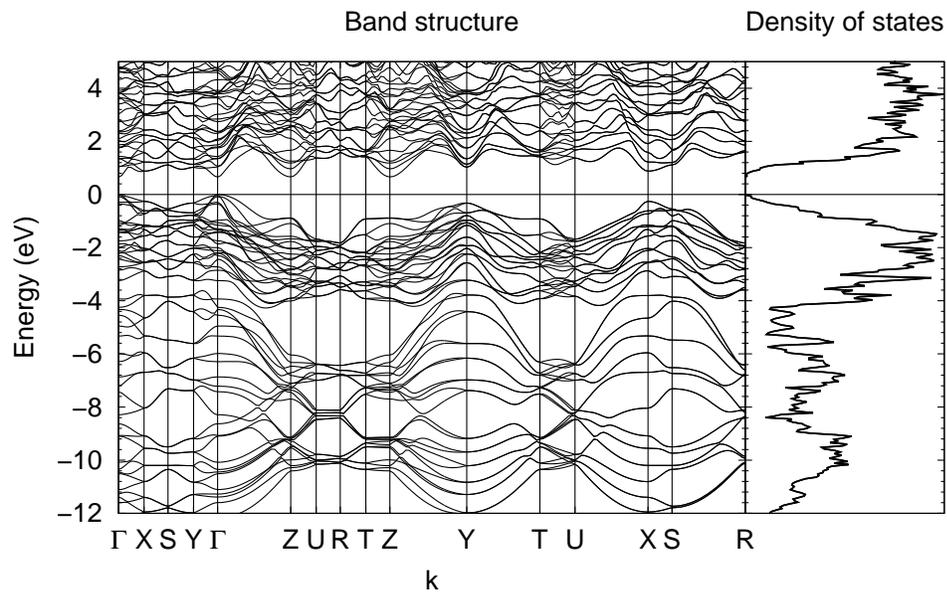}
\caption{Electronic band structure and DOS for germanium allotrope
Ge$\#28$.}\label{fig:BandGe}
\end{figure}

\end{document}